\documentclass[journal, twoside, final]{IEEEtran}
\usepackage{cite}
\usepackage{graphicx}
\usepackage{amssymb, amsmath, amsthm}
\usepackage{multirow}
\usepackage{color}
\usepackage{url}
\usepackage{setspace}
\usepackage{placeins}
\usepackage{booktabs, multicol, multirow}
\usepackage{acronym}
\usepackage{bm}
\usepackage{placeins}
\usepackage{extarrows}
\usepackage{enumitem}
\usepackage{verbatim}
\allowdisplaybreaks
\newtheorem{lemma}{Lemma}
\newtheorem{theorem}{Theorem}

\newtheorem{corollary}{Corollary}

\usepackage{soul}
\usepackage{siunitx}

\usepackage[caption=false]{subfig}
\usepackage{adjustbox}

\usepackage{algorithm, algorithmic}

\newmuskip\pFqmuskip
\newcommand*\pFq[6][8]{%
  \begingroup 
  \pFqmuskip=#1mu\relax
  \mathcode`\,=\string"8000
  \begingroup\lccode`\~=`\,
  \lowercase{\endgroup\let~}\pFqcomma
  {}_{#2}F_{#3}{\left[\genfrac..{0pt}{}{#4}{#5};#6\right]}%
  \endgroup
}
\newcommand{\pFqcomma}{\mskip\pFqmuskip}

\usepackage{etoolbox}
\usepackage{tikz}
\usetikzlibrary{shapes,backgrounds}

\newcommand{\tstar}[5]{
\pgfmathsetmacro{\starangle}{360/#3}
\draw[#5] (#4:#1)
\foreach \x in {1,...,#3}
{-- (#4+\x*\starangle-\starangle/2:#2) -- (#4+\x*\starangle:#1)
}
-- cycle;
}

\newrobustcmd*{\mystar}[1]{\tikz{\tstar{0.05}{0.1}{5}{56}{ultra thin, color=red, fill=red };}}

\newrobustcmd*{\mycircle}[1]{\tikz{\filldraw[draw=#1, fill=#1] (0,0) circle [radius=0.1cm];}}

\newrobustcmd*{\mytriangle}[1]{\tikz{\filldraw[draw=#1, fill=#1] (0,0) -- (0.2cm,0) -- (0.1cm,0.2cm);}}

\begin{document}

\title{Air-to-Ground Communications Beyond 5G: \\ CoMP Handoff Management in UAV Network}
	\author{Yan Li, Deke~Guo, Lailong Luo, and Minghua~Xia, \IEEEmembership{Senior Member, IEEE}
\thanks{Received 29 February 2024; revised 19 August 2024; accepted 5 October 2024. This work was supported in part by the National Natural Science Foundation of China under Grant U23B2004, Grant 62171486, and Grant U2001213; in part by the China Postdoctoral Science Foundation under Grant 2023M734314 and Grant 2024T71181; in part by the Postdoctoral Fellowship Program under Grant GZC20233529; and in part by the Guangdong Basic and Applied Basic Research Project under Grant 2021B1515120067. The associate editor coordinating the review of this paper and approving it for publication was Z. Guan. \itshape{(Corresponding authors: Deke~Guo; Minghua~Xia.)}}
\thanks{Yan Li is with the National Key Laboratory of Information Systems Engineering, National University of Defense Technology, Changsha, 410073, China, and also with the School of Computer and Communication Engineering, Changsha University of Science and Technology, Changsha, China (e-mail: liyanylnudt@nudt.edu.cn).
} 
\thanks{Deke Guo and Lailong Luo are with the National Key Laboratory of Information Systems Engineering, National University of Defense Technology, Changsha, 410073, China (e-mail: \{dekeguo, luolailong09\}@nudt.edu.cn). 
} 
\thanks{Minghua Xia is with the School of Electronics and Information Technology, Sun Yat-sen University, Guangzhou 510006, China (e-mail: xiamingh@mail.sysu.edu.cn).}
	\thanks{%
	Color versions of one or more of the figures in this article are available at https://doi.org/10.1109/TWC.2024.3476943.
	
	Digital Object Identifier 10.1109/TWC.2024.3476943}
}

\maketitle

\IEEEpubid{\begin{minipage}{\textwidth} \ \\[12pt] \centering 1536-1276 \copyright\ 2024 IEEE. Personal use is permitted, but republication/redistribution requires IEEE permission. \\
See \url{https://www.ieee.org/publications/rights/index.html} for more information.\end{minipage}}

\begin{abstract}
\noindent
Air-to-ground (A2G) networks, using unmanned aerial vehicles (UAVs) as base stations to serve terrestrial user equipments (UEs), are promising for extending the spatial coverage capability in future communication systems. Coordinated transmission among multiple UAVs significantly improves network coverage and throughput compared to a single UAV transmission. However, implementing coordinated multi-point (CoMP) transmission for UAV mobility requires complex cooperation procedures, regardless of the handoff mechanism involved. This paper designs a novel CoMP transmission strategy that enables terrestrial UEs to achieve reliable and seamless connections with mobile UAVs. Specifically, a computationally efficient CoMP transmission method based on the theory of Poisson-Delaunay triangulation is developed, where an efficient subdivision search strategy for a CoMP UAV set is designed to minimize search overhead by a divide-and-conquer approach. For concrete performance evaluation, the cooperative handoff probability of the typical UE is analyzed, and the coverage probability with handoffs is derived. Simulation results demonstrate that the proposed scheme outperforms the conventional Voronoi scheme with the nearest serving UAV regarding coverage probabilities with handoffs. Moreover, each UE has a fixed and unique serving UAV set to avoid real-time dynamic UAV searching and achieve effective load balancing, significantly reducing system resource costs and enhancing network coverage performance.
\end{abstract}

\acresetall  


\acrodef{3GPP}{}
\acrodef{CoMP}{}
\acrodef{LTE}{Long-Term Evolution}
\acrodef{UE}{user equipment}

\IEEEpubidadjcol

\section{Introduction}
\label{Section:Introduction}
To meet the demands of beyond 5G space-air-ground integrated network interconnects, ubiquitous wireless connectivity with high data rates and high-reliability requirements becomes imperative \cite{10054381}. Despite the widespread adoption of conventional terrestrial communication systems, a significant portion of the worldwide population still needs to be connected or more connected due to the limitation of terrestrial infrastructure \cite{9042251}. Remarkably, there is an urgent need to enhance broadband coverage in remote areas\cite{9275613}. In this regard, air-to-ground (A2G) communications are vital in extending telecommunication coverage and providing emergency network recovery \cite{8579209}. Meanwhile, coordinated multi-point (CoMP) transmission \cite{3GPPTR36.819}, a collaborative technique employed among multiple base stations (BSs) in legacy cellular networks, demonstrates exceptional effectiveness in mitigating inter-cell interference (ICI) and improving network coverage and throughput. By incorporating CoMP transmission technologies, in recent years, the A2G network has become a promising paradigm to support high data rates and high-reliability transmission services, especially in suburban, rural, or remote areas. However, it also introduces complex cooperation strategies and handoff patterns that may degrade the performance of terrestrial user equipments (UEs). In this paper, we develop a mathematically tractable A2G communication scheme and devise effective subdivision search and cooperative handoff management strategies.

\IEEEpubidadjcol

\subsection{Related Works and Motivations}
\label{Section_Relatedwirks}
In general, current wireless network models primarily consist of two categories: random network models and stochastic geometric models. In principle, they differ in their approach to understanding network behaviors: The former focuses on the randomness of connections between network entities, while the latter concentrates on the stochastic geographic locations of network entities.

In the random network models, BSs are deployed using convex optimization and machine learning techniques to optimize system performance and achieve satisfactory performance metrics for target UEs. For instance, the work \cite{8038869} deployed multiple unmanned aerial vehicles (UAVs) for energy-efficient data collection in the Internet of Things by using the convex optimization theory. The study \cite{8038014} designed an optimal three-dimensional (3D) deployment strategy for UAVs to maximize the number of UEs while focusing on the quality of service. To provide large-scale network connections for terrestrial UEs, a multi-UAV auxiliary design and optimization scheme was proposed in \cite{8624565}. On the other hand, machine learning theory has been widely employed in recent years to characterize the performance of A2G networks. For instance, an adaptive and efficient deployment strategy for a UAV-assisted network based on mixed deep reinforcement learning was proposed in \cite{9348512}, and this strategy enables UAVs to adjust their movement direction and distance based on the UE's random movement trajectory. 

In the stochastic geometric models, the point process theory was widely applied to characterize the network performance in finite and infinite regions \cite{Haenggi12}. For instance, a 3D modeling technique using a two-dimensional (2D) labeled Poisson point process (PPP) was developed in \cite{9332278}. This method takes each network node and associated tags as the distinctive ground projection and random height in the pertaining 3D point process. In \cite{9577015}, a spatial PPP was employed to model the positions of UAVs, where the effect of UAV altitude and density on the coverage probability and average achievable data rate for terrestrial UEs was analyzed. To investigate the high-order statistics of the signal-to-interference ratio (SIR) of UEs and distinguish between single links, a comprehensive analytical framework based on the Meta distribution was proposed in  \cite{10003212} for A2G networks. Furthermore, for scenarios with a limited number of nodes, the binomial point process (BPP) was employed to simulate UAV locations, followed by an analysis of downlink coverage performance \cite{8671460}. Analogously, the randomness of UAV BSs in confined spaces was studied in \cite{8525328}, yielding a tractable probability distribution function for the SIR of the typical UE. Also, the coverage probability of a truncated octahedron-based 3D UAV network was derived in \cite{9740446}. On the other hand, some advanced point process theory was applied to deal with potential correlations between network nodes. For instance, the Poisson hole process (PHP) was applied in \cite{9667269} to model UAV nodes and derive optimal deployment parameters for maximum network coverage. The spatial hard-core point process (HCPP) was employed to model a UAV network in \cite{9831248}. In this paper, the Poisson point process is used to model the random positions of UAVs, and then the Poisson-Delaunay triangulation is used to model the UAV swarm network.

To enhance spectral efficiency, especially in cell-edge areas, and mitigate ICI, CoMP transmission technology has emerged as a practical and viable solution by allowing UEs to communicate with multiple nearby BSs in a specially coordinated manner. In industry, the way to implement CoMP was initially described in 3GPP TR 36.819, and some features were further enhanced in 3GPP TR 36.741. In the context of A2G networks, a multi-agent decentralized dual-depth Q method was proposed in \cite{9757149} to optimize the UAV energy efficiency. To minimize the weighted energy consumption of UAVs, a multi-agent cooperation approach based on multi-agent proximal policy optimization was developed in \cite{10021296}. To jointly schedule sensing, computing, and communication resources, a multi-agent reinforcement learning-based algorithm for UAS swarms was designed in \cite{9802876}. The maximum task completion delay was minimized by using UAVs to cooperatively provide computing and caching resources for ground devices in \cite{9798889}. The cooperative Internet of UAVs was studied in \cite{9154432}, addressing how UAVs accomplish perceptual tasks through CoMP to minimize the impact of the information age. Meanwhile, a multi-UAV cooperative sensing and transmission scheme was proposed in \cite{9968197} to reduce the overall mission completion time. Like the typical hexagonal cell in terrestrial networks, a binomial-Delaunay tetrahedralization was devised as a typical cell in \cite{9151343}, allowing four coordinated UAVs to serve UEs jointly. Nevertheless, the papers mentioned above ignored the searching strategy for a UAV CoMP set and the UAV mobility to ease mathematical tractability.
 
In general, the UAV mobility models in the open literature can be categorized into tracking-based ones and stochastic synthesis ones \cite{8673556}. In particular, the stochastic synthesis models were widely applied for analyzing and optimizing various network scenarios, thanks to their adaptability and high mathematical tractability. For instance, a hybrid UAV mobility model based on a random waypoint model was proposed in \cite{8421028}, assuming a fixed distance between the typical UE and its serving UAV. A modified random direction mobility model was developed in \cite{8533634} to simulate the mobile characteristics of UAVs. Following the UAV simulation models specified in 3GPP TR 36.777, a simplified stochastic model was implemented in \cite{9013645}, where the UAV moves in a straight line at a constant speed and a fixed height.

In light of UAV mobility, the handoff performance becomes crucial for the efficiency and reliability of A2G networks. In general, the analytical methods to evaluate the handoff performance can be categorized into trajectory-based methods and distance correlation-based methods \cite{8673556}.  For instance, the user handoff rate was derived based on Voronoi cells in  \cite{6477064} by counting the number of times the user's movement intersected the cell boundaries. On the other hand, the handoff probability for the typical UE was analytically derived in \cite{7006787} by defining the likelihood of switching to another cell in a mobile cycle based on the nearest distance criterion. In 3D space, the equivalence of vertical and horizontal handoffs was studied in \cite{7866856}. Using CoMP, a cooperative handoff management scheme was designed in \cite{7841695} to reduce the effect of handoff on throughput gains achieved through network densification. Regarding spatial scenarios, the cooperative handoff performance of UAVs in a traditional hexagonal grid was analyzed in \cite{8998329}. However, the hexagonal cell model does not effectively represent the actual deployment of spatial nodes. To the best of the authors' knowledge, the analytical characterization of handoff probability in A2G networks is still an open problem.

\subsection{Major Contributions}
This paper first develops a mathematically tractable A2G network model based on CoMP transmission. Then, a subdivision search strategy is designed to effectively minimize the search overhead for a cooperating UAV set. Finally, the dynamic cooperative handoff and coverage probabilities for the typical terrestrial UE are analytically derived. In summary, the major contributions of this paper are threefold:
\begin{enumerate}[label = {\arabic*)}]
	\item  Network modeling: A novel A2G network model is proposed based on the theory of Poisson-Delaunay triangulation, which ingeniously integrates the CoMP technology to enhance the network coverage. In particular, each UE is jointly served by three UAVs shaping a triangular cell, yielding a deterministic CoMP set of UAVs without exhaustive searching.

	\item CoMP set search strategy: By the relation between the network node density and the node degrees of freedom (DoF), the minimum search area for a CoMP set is constructed. Then, a dynamic subdivision search strategy using a recursive divide-and-conquer method is designed to minimize the search overhead for a cooperating UAV~set.
	
	\item  Cooperative handoff management and coverage probability analysis: A simple yet effective handoff mechanism is conceived using the equivalence between the Poisson-Delaunay triangulation and its dual Voronoi tessellation. Then, lower bounds on the cooperative handoff probability and the coverage probability with handoffs are derived for the typical UE. 
\end{enumerate}

\subsection{Paper Organization}
The remainder of this paper is organized as follows: Section \ref{Section_SystemModel} describes the system
model. Section~\ref{Section_CoMPSet} develops a subdivision search strategy for a UAV CoMP set. Section~\ref{Section_Cov_Prob} analyzes the performance of the typical UE regarding the handoff probability and the coverage probability with handoffs. Simulation and numerical results are extensively presented and discussed in Section~\ref{Section_Simulation}. Finally, Section~\ref{Section_Conc} concludes the paper.

{\it Notation}: Scalars, vectors, and matrices are denoted by italic letters and lower- and uppercase bold letters, respectively. The operator ${\rm round}(\cdot)$ yields the nearest integer to a real number, and $\lfloor \cdot \rfloor$ denotes the largest integer less than or equal to a real number. The operators $\mathbb{E}( \cdot )$ and ${\rm Var}( \cdot )$ computes the mathematical expectation and variance of a random variable, respectively. The notation $\|\bm{x}\|$ and $\bm{x}^H$ compute the $\ell_2$-norm and Hermitian transpose of vector~$\bm{x}$, respectively. The symbol $\Gamma(a, b)$ represents a Gamma distribution with the shape parameter $a$ and the scale factor $b$. The Gamma function is defined as $\Gamma(c) \triangleq \int_0^\infty{x^{c-1} \exp(-x)}\,{\rm d}x$, for all $c > 0$. The generalized hypergeometric function is defined as ${\displaystyle \, {}_{m}F_{n}(a_{1},\cdots ,a_{m}; b_{1}, \cdots, b_{n}; x) \triangleq \sum_{p=0}^{\infty }{\frac {(a_{1})_{p} \cdots (a_{m})_{p}}{(b_{1})_{p} \cdots (b_{n})_{p}}} \, {\frac {x^{p}}{p!}}}$, with $(b)_p = b(b+1)\cdots(b+p-1)$ if $p > 1$ and $(b)_p = 1$ if $p = 0$.

\section{System Model}
\label{Section_SystemModel}
\begin{figure}[!t]
	\centering
	\includegraphics [width=3.25in, clip]{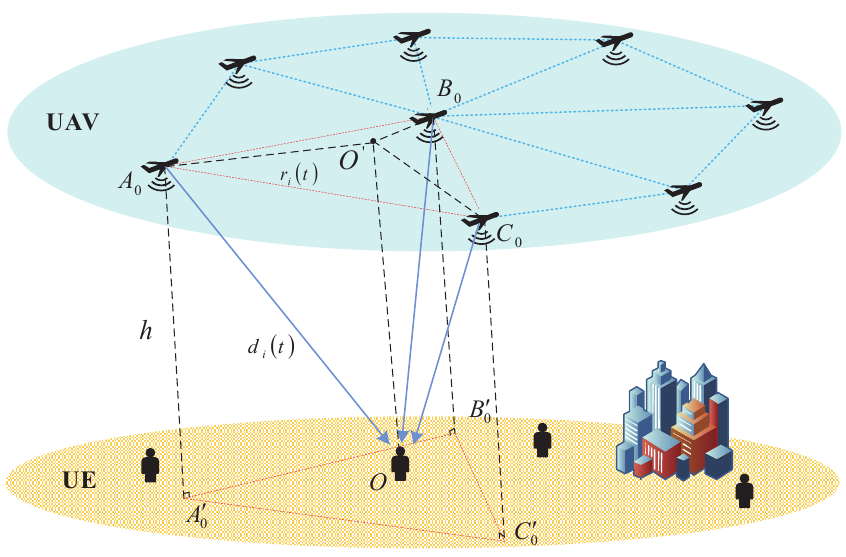}
	\caption{The proposed A2G network based on the Poisson-Delaunay triangulation, where the typical terrestrial UE is served by the UAVs in the CoMP set $\{A_0, B_0, C_0\}$ with distances $d_i$, $i = 1, 2, 3$.}
	\label{Fig-1}
\end{figure}

As shown in Fig.~\ref{Fig-1}, this paper studies an A2G network where multiple UAVs equipped with aerial BSs serve numerous terrestrial UEs.  Each UAV is assumed to be equipped with $M$ antennas and hovers at the same height $h$\footnote{In real-world applications, different UAVs generally have diverse flight heights. Still, their height difference is usually negligible compared to the distance between them and terrestrial UEs. \label{Footnote-1}}, and each UE is supposed to have a single antenna. At the initial time with $t = 0$, all UAVs are assumed to be distributed as a homogeneous PPP, denoted $\Phi(0)$, with intensity $\lambda_0$. Without loss of generality, the typical UE, independent of the UAV distribution, is supposed to be located at the origin on the ground, say, $O(0, 0, 0)$. To enhance the quality of service of terrestrial UEs, the CoMP transmission technique is employed, and the typical UE is served by three UAVs in the CoMP set, say,  $U(0) = \{A_0, B_0, C_0\}$. The projection of $O$ on the ground onto the UAV plane is denoted $O^{\prime}$. To account for the UAV mobility in the case of $t > 0$, the evolution of UAV positions over time $t$ is subject to a point process, denoted $\Phi(t)$, with intensity $\lambda(t)$. Accordingly, the distances of a UAV within the corresponding CoMP set $U_0(t)$ from the projected $O'$ and the origin $O$ on the ground are denoted by $r_i(t)$ and $d_i(t) = (r^2_i(t) + {h}^2)^{1/2}$, $\forall i \in \{1, 2, 3\}$, respectively. Also, the distance between the nearest UAV to the projected $O^{\prime}$ is counted by $r^{*}(t) \triangleq \min\{r_i(t), i = 1, 2, 3\}$.

\subsection{The CoMP Set of a UE}
\label{The CoMP Set of a UE}
\begin{figure}[!t]
	\centering
	\includegraphics [width=3.5in, clip]{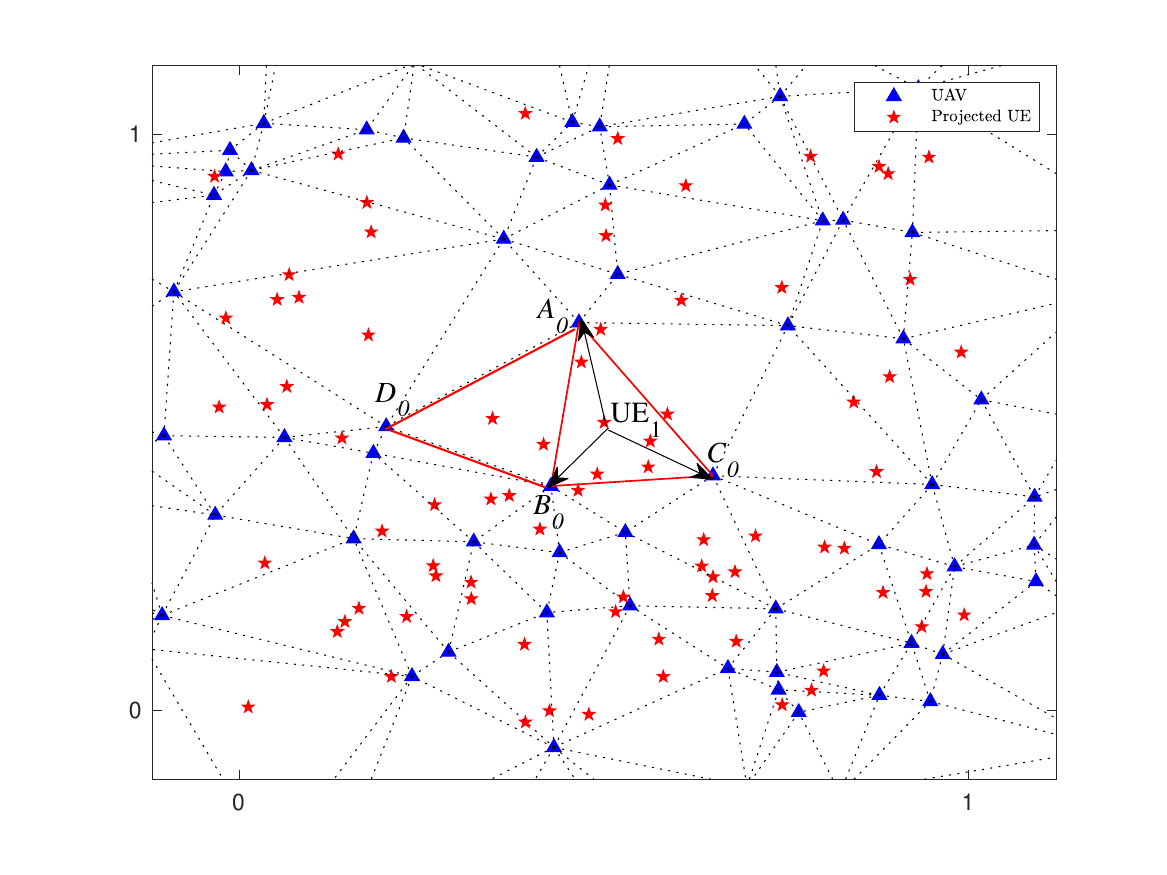}
	\vspace{-20pt}
	\caption{An illustrated Poisson-Delaunay triangulation network where CoMP sets consist of the UAVs at the vertices of Delaunay triangles with blue dash boundaries. The projected UAVs (denoted by blue solid triangles) are distributed as a PPP, and the UEs (denoted by red stars) are uniformly distributed in a normalized coverage area of one square kilometer.}
	\label{Fig_2}
\end{figure}

To simplify the 3D scenario previously sketched in Fig.~\ref{Fig-1}, Fig.~\ref{Fig_2} illustrates UAVs and the projected UEs on the same 2D plane, denoted by blue solid triangles (`\mytriangle{blue}') and red stars (`\mystar{red}'), respectively. If each projected UE is associated with its nearest UAV, the resulting polygonal boundaries define a Poisson-Voronoi tessellation, like that in legacy cellular networks. In contrast, the dual Poisson-Delaunay triangulation is plotted in Fig.~\ref{Fig_2} as the triangles with blue dash boundaries. For more details about the relation between Poisson-Voronoi tessellation and its dual Poisson-Delaunay triangulation, the interested reader is referred to \cite{8358978, 8976426} and references therein. 

For each projected UE, the corresponding CoMP set can be constructed as follows. By considering the projected ${\rm UE}_1$ in Fig.~\ref{Fig_2} for instance, it initially chooses the nearest UAVs $A_0$ and $B_0$ as its two associated BSs, thereby establishing an edge of a cooperative triangular cell. Two adjacent triangles sharing the same edge define a quadrilateral with vertices $\{A_0, B_0, C_0, D_0\}$. From the four UAVs at the quadrilateral vertices, ${\rm UE}_1$ selects the two nearest UAVs and another UAV located between the remaining two opposing UAVs, but closer to ${\rm UE}_1$, to finally constitute the CoMP set $U(0) = \{A_0, B_0, C_0\}$. For the typical UE, the UAVs within the CoMP set are designated as serving UAVs, while all the other UAVs are treated as interfering ones.

\subsection{Mobility Model}
To adapt the UAV mobility model accepted by the 3GPP \cite{Rep36.777, 9013645}, we consider the different moving speeds (DMS) model in this paper. Specifically, each UAV is supposed to move at a random velocity and height along a straight line but in a uniformly distributed random direction, independent of any other UAV. Indeed, the DMS model is a generalization of the same moving speed (SMS) model, wherein UAVs move at the same speed in a uniformly random direction. Albeit simple, the DMS mobility model can be taken as a benchmark for evaluating more sophisticated mobility models. 

\begin{figure}[!t]
	\centering
	\includegraphics [width=2.25in, clip]{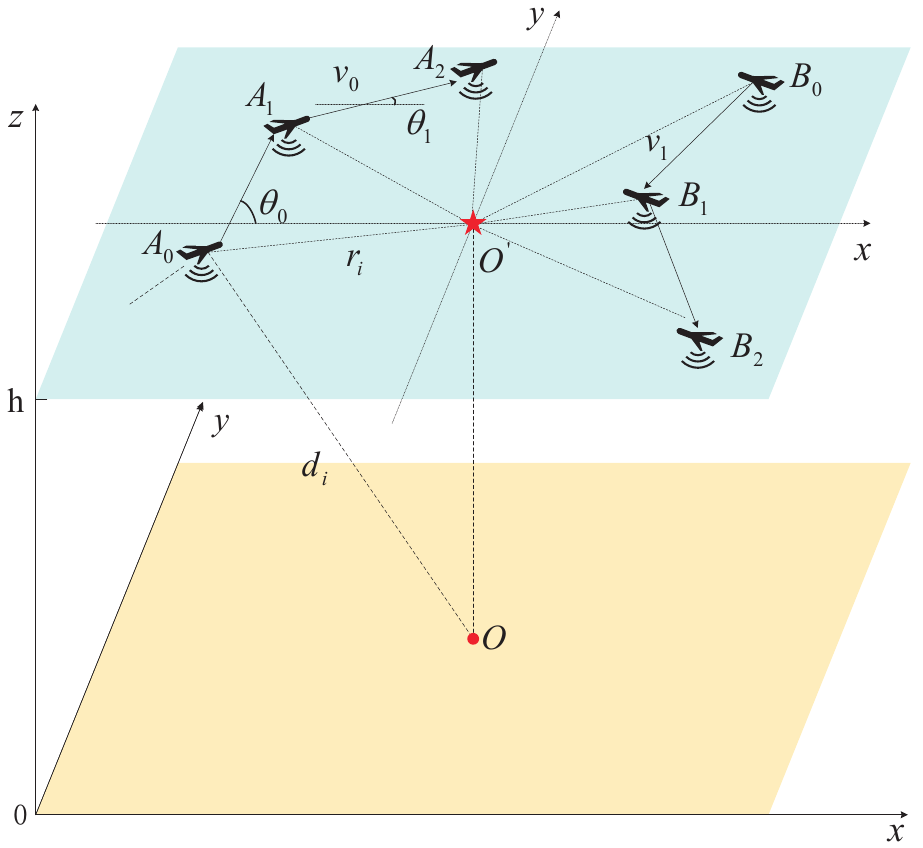}
	\vspace{-10pt}
	\caption{An illustration of the different moving speed (DMS) mobility model.}
	\label{Fig_3}
\end{figure}

For illustrative purposes, Fig.~\ref{Fig_3} plots the moving trajectories of two UAVs with independent and identically distributed (i.i.d.) random speeds $v_0$ and $v_1$, in which  $A_i$ and $B_i$ represent their turning points at time $t_i$, $i \in \{0, 1, 2\}$, respectively. Also, the angle between the direction $A_0 \to A_1$ and the positive direction of the $x$-axis is denoted by $\theta_0$, which is supposed to be uniformly distributed in $[0, 2\pi)$. When the UAV reaches $A_1$, it changes with a new direction of angle $\theta_1$ to reach the next turning point $A_2$, where $\theta_1$ has independent and identical distribution to $\theta_0$.

\subsection{Channel Model}
For ease of mathematical tractability and capturing a broad range of fading scenarios, we employ the Nakagami-$m$ fading model in this paper, and its relationship to the more general $\eta-\mu$ model will be discussed later. All UAVs are assumed to have the same transmit power $P$, then the received signal power at the typical UE from the three cooperative UAVs in the set $U_0(t) = \{A_0, B_0, C_0\}$, can be computed as 
\begin{equation} \label{Eq-Rx}
	S(t) = \left|\sum_{i \in U_0(t)} \hspace{-0.5em} P^{\frac{1}{2}} d^{-\frac{\alpha}{2}}_i(t) \bm{g}^{H}_i(t) \bm{\omega}_i(t) \right|^2, 
\end{equation}
where $d_i(t)$ indicates the Euclidean distance from the $i^{{\rm th}}$ UAV to  the typical UE (cf. Fig.~\ref{Fig-1}); $\bm{g}_i(t) \in \mathbb{C}^{M \times 1}$ is the complex-valued channel vector from the $i^{\mathrm {th}}$ serving UAV to  the typical UE; $\bm{\omega}_i(t) \in \mathbb{C}^{M \times 1}$ denotes the pre-coder used at the $i^{{\rm th}}$ UAV, and $\alpha > 2$ refers to the path-loss exponent (PLE). Also, suppose full downlink channel state information (CSI) is available at UAVs, then the channel-inverse pre-coder $\bm{\omega}_i(t)$ used by the $i^{{\rm th}}$ BS is given by $\bm{\omega}_i(t) = {\bm{g}_i(t)}/{\|\bm{g}_i(t)\|}$. Likewise, assuming all UAVs share the same spectrum resources, the interference power can be accounted for as
\begin{equation} \label{Eq-Interference}
	I(t) = \hspace{-1em} \sum_{j \in \Phi(t) \setminus U_0(t)} \hspace{-1.5em}  P d^{-\alpha}_{j,0}(t)h_j(t),
\end{equation}
where $h_j(t) \triangleq |\bm{g}^{H}_{j,0}(t)\bm{\omega}_j(t)|^2$; the difference set $\Phi(t) \setminus U_0(t)$ represents the point process of interfering UAVs, and $d_{j,0}(t)$ and $\bm{g}_{j,0}(t)$ denote the Euclidean distance and the complex-valued channel vector from the $j^{{\rm th}}$ interfering UAV to the typical UE, respectively. Notice that intra-cell interference is not accounted for in \eqref{Eq-Interference} because it can be effectively mitigated in practice by using techniques such as successive interference cancellation or orthogonal frequency division multiple access.

\subsection{Performance Metrics}
In light of \eqref{Eq-Rx}-\eqref{Eq-Interference} and ignoring the noise at UEs, the received SIR at the typical UE can be expressed as
\begin{equation}\label{Eq_RxSINR}
	\Upsilon(t) \triangleq \frac{S(t)}{I(t)} = \frac{\bigg|\sum\limits_{i \in U_0(t)} \hspace{-0.5em} d^{-\frac{\alpha}{2}}_i(t) \bm{g}^{H}_i(t)\bm{\omega}_i(t)\bigg|^2}{\sum\limits_{j \in \Phi(t) \setminus U_0(t)} \hspace{-1.5em} d^{-\alpha}_{j,0}(t)h_j(t)}.
\end{equation}
Then, like \cite{9893878,7006787}, given a predetermined SIR threshold value $\gamma$, the coverage probability with handoffs at time $t$ is defined as
\begin{align}
	\hspace{-0.6em} \mathcal{P}(t) &\triangleq \Pr \left\{\Upsilon(t) > \gamma, {\rm \overline{Hf}} \right\} + (1-\zeta)\Pr\left\{ \Upsilon(t) > \gamma, {\rm Hf} \right\} \label{Eq_Coverage_Handoff_form1} \\
	&= (1-\zeta)\Pr \left\{\Upsilon(t) > \gamma \right\} + \zeta \Pr \left\{\Upsilon(t) > \gamma, {\rm \overline{Hf}} \right\} \nonumber \\
	&=\left[(1-\zeta)+ \zeta \left(1-\mathbb{P}[H(t)] \right) \right] \mathbb{P}_{\rm{C}}(t), \label{Eq_Coverage_Handoff_form2}
\end{align}
where the abbreviations ${\rm {Hf}}$ and ${\rm \overline{Hf}}$ in \eqref{Eq_Coverage_Handoff_form1} represent the events with and without a handoff, respectively. Also, the factor $\zeta \in [0, 1]$ denotes the probability of connection failure attributed to handoffs, indicating the system's tolerance towards handoffs. As a result, the first term on the right-hand side of \eqref{Eq_Coverage_Handoff_form1} denotes the probability of a UE being in coverage without a handoff. The second term represents the probability of a UE experiencing a handoff within coverage, adjusted by the handoff cost $1-\zeta$. In \eqref{Eq_Coverage_Handoff_form2}, $\mathbb{P}[H(t)] \triangleq \Pr\{\rm Hf \}$ indicates the handoff probability and $\mathbb{P}_{\rm C}(t) \triangleq \Pr \left\{\Upsilon(t) > \gamma \right\}$ refers to the coverage probability without any handoff, with the subscript `C' indicating `Coverage'. The two probabilities will be explicitly evaluated later.

\section{The Subdivision Search Strategy}
\label{Section_CoMPSet}
Due to UAV mobility, the status of network links undergoes continuous changes. Consequently, UEs must relocate or search for their CoMP sets periodically, and the search overhead escalates if there are more and more UAVs. Accordingly, in this section, we develop a practical subdivision search strategy to mitigate the searching overhead. In particular, we employ a divide-and-conquer strategy \cite[Ch. 4]{Hjelle06} and divide the entire UAV set into roughly equal-sized subsets. The UAV subsets consisting of the least number of UAVs are called the minimum search subsets. This section aims to determine a cooperative UAV set for the typical UE within a limited number of minimum search subsets.

\subsection{The Minimal Search Radius}
Before elaborating on the subdivision search scheme, we first determine the effective radius of the search region (say, $R_s$), which includes at least a complete cooperative UAV set. From the probabilistic perspective, we evaluate the probability of a circle with radius $R_s$ containing at least one UAV CoMP set by the geometric relation between the density $\lambda_0$ and the DoFs of UAVs. It is easily known that each UAV has six DoFs by recalling the theory of stochastic geometry \cite[Table 5.11.1]{OkabeBoots2000}. Therefore, an average minimum of $n = 18$ UAVs is necessary to establish a complete cooperative set, given that each CoMP set consists of three UAVs. Accordingly, given UAV density $\lambda_0$, the search area for UAVs should satisfy the inequality $	\lambda_0 \pi R_s^2 \geq 18$. As a result, the minimum search radius can be determined as
\begin{equation} \label{Eq_Rs_Minimal}
	R_{s \min} = \sqrt{\frac{18}{\lambda_0 \pi}}.
\end{equation}
Consequently, the UAVs within the minimum search region collectively constitute a subdivision set. When the UAV number reaches 18, the probability of containing a complete CoMP set is optimized for the lowest search cost. 

\begin{figure}[t!] 
	\centering
    \includegraphics [width=3.25in, clip]{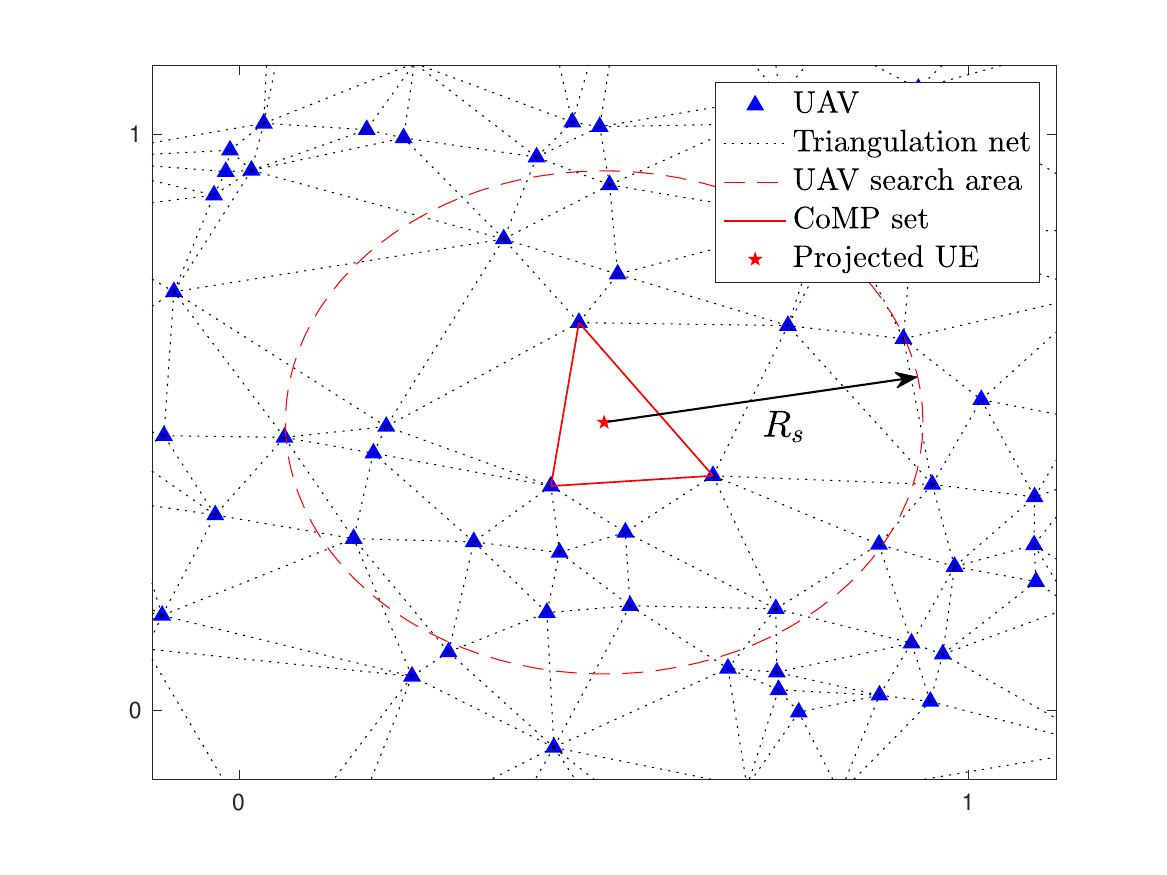}
	\caption{Illustration of a subdivision searching process for a CoMP set.} 
	\label{Fig-4}  
\end{figure}

\subsection{Subdivision Search Strategy For A CoMP Set }
Fig.~\ref{Fig-4} illustrates a finite plane of UAV intensity $\lambda_0$. We first construct the Delaunay triangulation using the divide-and-conquer algorithm \cite[Ch. 4]{Hjelle06}. Then, we calculate the radius of the search area, corresponding to the red dashed circle with radius $R_s$ shown in Fig.~\ref{Fig-4}. Once the search area is established, the CoMP set can be constructed as follows. Let $m$ be the total number of UAVs spread across the plane, and $n$ denote the number of UAVs within the minimum search circle. The distance of a UAV located at $\bm {x}_k(t)$ at time $t$ from $O$ is denoted by $d_k(t)$, $1 \leq k \leq m$. The locations $\bm{x}'_i(t)$ of UAVs within the minimum search area can then be determined by $d_k(t) \leq R_{s \min}$, which consist of the set $\bm{B}(t)=\{\bm{x}'_1(t), \bm{x}'_2(t), \cdots, \bm{x}'_n(t)\}$. Therefore, the Delaunay triangular connection set $\bm{L}_i(t) \triangleq \{L_1(t), L_2(t), \cdots, L_i(t)\}$ are determined according to $\bm{B}(t)$ with $1\leq i \leq I$, where $I$ denotes the cardinality of $\bm{L}_i(t)$ and each triangular connection list $L_i(t)$ consists of three UAVs. Finally, as shown by the red triangle in Fig.~\ref{Fig-4}, the CoMP set $U_0(t)$ serving the projected UE consists of the three UAVs with the strongest average received power, where the average is taken with respect to the fading coefficient. Mathematically, the CoMP set is determined by
\begin{equation} \label{Eq_CoMP_Set}
	U_0(t) = \mathop{\arg \rm{max}}\limits_{\{x'_1(t), x'_2(t), x'_3(t)\} \subset \cup_{j=1}^{I}L_j(t)} \sum\limits_{i=1}^3 \|\bm{x}'_i(t)\|^{-\frac{\alpha}{2}}. 
\end{equation}

In summary, the proposed subdivision search scheme for a CoMP set is formalized in Algorithm~\ref{alg_subdivision}. Incidentally, the subdivision search strategy given by \eqref{Eq_CoMP_Set} can be easily extended to more general cases where UAVs fly at varying heights. 

\begin{algorithm} [!t]
	\small
	\caption{The Proposed Subdivision Search Scheme}
	\label{alg_subdivision}
	\begin{algorithmic}[1]  
		\REQUIRE  The density of UAV $\lambda_0$,  the PLE $\alpha$, and the height of UAVs $h$.
		\STATE {Construct the Delaunay triangulation;}
		\STATE {Compute the distance $d_k(t)$ between a UAV located at $\bm {x}_k(t)$ at time $t$ and the origin $O$, $1\leq k \leq m$, where $m$ denotes the total number of UAVs;}
		\STATE {Compute the search radius $R_{s \min}$ as per \eqref{Eq_Rs_Minimal};}
		\STATE {For the search circle, determine the triangular connection set $\bm{L}_i(t)$ at time $t$, $i = 1, \cdots, I$;}
		\FOR {$i := 1$ to $I$ \do} 
		\STATE {Compute the average received power for each triangular list $L_i(t)$ in the minimum search circle;}
		\ENDFOR 
		\STATE {Determine the CoMP set of UAVs according to \eqref{Eq_CoMP_Set}.}
	\end{algorithmic}
\end{algorithm}

Finally, we evaluate the computational complexity of Algorithm~\ref{alg_subdivision}. Given the UAV density $\lambda_0$, the average number of UAV nodes within the search circle is  $\lambda_0\pi R^2_s$. By the computational complexity of the optimal triangulation construction algorithm as detailed in \cite[Thm. 4.3]{Hjelle06}, the proposed Algorithm~\ref{alg_subdivision} exhibits a computational complexity of $O\left(\lambda_0\pi R^2_s \log{\left(\lambda_0\pi R^2_s\right)}\right)$.

\section{CoMP Handoff Strategy and Performance Evaluation}
\label{Section_Cov_Prob}
To adapt UAV mobility, an equivalent Voronoi approximation method is designed in this section to characterize the dynamic handoff area of the typical UE. First, we characterize the density of the non-serving UAVs; then, we derive the speed and direction of the CoMP UAV set movement. Further, a lower bound on the dynamic cooperative handoff probability is developed. An integral expression on the handoff probability for the SMS model is also derived for comparison purposes. Finally, a lower bound on the coverage probability with handoffs is established. 

\subsection{CoMP Handoff Strategy}
We first define a handoff operation occurring in our handoff management scheme. As shown in Fig.~\ref{Fig_2}, whether the typical UE performs a handoff depends on whether its average received power is changed or not, where the average is taken regarding channel fading. Let $U_j (t)= \{A_j, B_j, C_j\}$, $j = 0,1,\cdots$, denote the $j^{\rm th}$ CoMP UAV set, then we have $\Phi(t)= \{U_j(t)\}|_{j=0}^{\infty}$. If $E_j(t) = |\sum_{i \in U_j(t)} P^{\frac{1}{2}}(t) d^{-\frac{\alpha}{2}}_i(t)|^2 < E_k(s) = |\sum_{i \in U_k(s)} P^{\frac{1}{2}}(s) d^{-\frac{\alpha}{2}}_{i}(s)|^2$, then a handoff from the $j^{\rm th}$ CoMP UAV set to its adjacent $k^{\rm th}$ one occurs. 

\begin{figure}[!t]
	\centering
	\includegraphics [width=3.0in, clip]{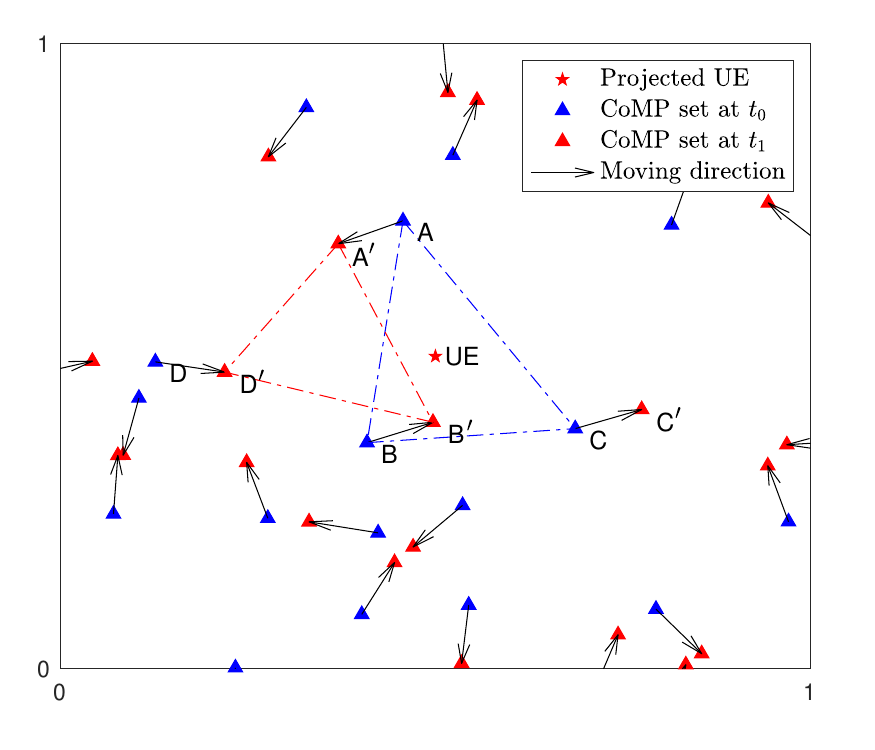}
	\vspace{-10pt}
	\caption{Illustration of the handoff from CoMP set $\{A, B, C \}$ to $\{A^{\prime}, B^{\prime}, D^{\prime}\}$ due to UAV mobility.}
	\label{Fig-5}
\end{figure}

For illustrative purposes, Fig.~\ref{Fig-5} illustrates the handoff process of CoMP UAV sets for the typical projected UE. The initial CoMP UAV set at time $t_0$ is denoted by $\{A, B, C\}$, and these UAVs turns to $\{A^{\prime}, B^{\prime}, C^{\prime}\}$ at the next time $t_1$ due to their independent movement. As the UAVs $\{A^{\prime}, B^{\prime}, C^{\prime}\}$ do not form a triangular Delaunay cell, the typical projected UE has to switch to its nearest CoMP UAV set $\{A^{\prime}, B^{\prime}, D^{\prime}\}$. Obviously, the location information and the structure of the triangular cells of the UAVs within a CoMP set for the typical UE are time-varying due to UAV mobility. As a result, the serving CoMP UAV set forming the associated triangular cell changes along with the time-varying Voronoi triangulation, requiring a handoff whenever the CoMP UAV set for the typical UE is changed. Therefore, it is challenging to characterize the coverage area of a virtual triangular cell, namely, the geographic region where a UE is associated with the same CoMP UAV set.  

\subsection{CoMP Handoff Probability}
Now, we analyze the handoff probability regarding the DMS mobility model. We also derive the CoMP handoff probability for the SMS model for comparison purposes. Before deriving the CoMP handoff probability, we first compute the density of interfering UAVs. 

As the nearest-neighbor association criterion introduces an exclusion zone, $\mathbb{C}=c(O,r^{*})$, the interfering UAVs follow an inhomogeneous point process. Taking $\mathbb{C}$ into consideration, the resulting UAV density can be divided into two exclusive parts: One part is contributed by $\mathbb{C}$, denoted $\lambda_1 (t; r_i, r^{*})$, i.e., due to the UAVs that are initially inside $\mathbb{C}$. The other part refers to the interfering UAVs, denoted $\lambda(t; r_i, r^{*})$, i.e., due to the points that are initially outside $\mathbb{C}$. Since the UAV density is $\lambda_0$ in total, then we get $\lambda(t; r_i, r^{*}) = \lambda_0-\lambda_1(t; r_i, r^{*})$. Applying an approach similar to \cite[Lemma~2]{9078878}, the density of interfering UAVs can be explicitly computed, as summarized in the following lemma.

\begin{lemma}\label{Lemma_DMS_Lambda}
	Let $v$ be a non-negative random variable denoting the speed of a UAV, with the cumulative density function (CDF) $F_v(x)$ and probability density function (PDF) $f_v (x)$, then the difference set $\Phi(t)\setminus U_0(t)$ forms an inhomogeneous PPP with intensity computed by
	\begin{align}\label{Lambda_DiffeSpeed}
		\lambda(t; r_i, r^{*}) 
		&= \lambda_0 - \lambda_0F_v\left(\frac{r^{*}-r_i}{t}\right) - \frac{\lambda_0}{\pi}\int_{\frac{|r^{*}-r|}{t}}^{\frac{r^{*}+r_i}{t}} f_v(x) \nonumber\\
		&\quad \times \arccos\left(\frac{{r_i}^2+v^2t^2-{r^{*}}^2}{2r_i vt}\right) {\rm{d}}x.
	\end{align} 
\end{lemma}

With \eqref{Lambda_DiffeSpeed}, we are in a position to derive the handoff probability for both the DMS and SMS models.

\subsubsection{\underline{Handoff Probability Regarding the DMS Model}}
We exploit an equivalent handoff model most recently developed in \cite{9893878} for ground-to-air wireless networks to address the intricate handoff patterns in a mobile UAV network. This model studies the handoff performance of a mobile UAV terminal served by multiple fixed terrestrial BSs. In analogy to the ground-to-air network, the average coverage area of the virtual Delaunay triangular cells in this paper is equivalent to that of the dual Voronoi cells since the coverage area of the triangular cells depends only on the density of the underlying PPP. Therefore, the CoMP handoff region, which was initially challenging to analyze, reduces to the classical Voronoi structure with the nearest neighbor association criterion. The resulting Voronoi cells exhibit tractable convexity, allowing for their quantitative characterization by stochastic geometry theory. However, to deal with UAV mobility, the velocity (including the speed and direction components) of the equivalent serving UAVs must be computed, as detailed below.
 
Without loss of generality, we assume that the typical UE receives signals from the three serving UAVs in the CoMP set $U_0(t) = \{A_0, B_0, C_0\}$, and each UAV in $U_0(t)$ moves at speed $v_i$ in direction $\theta_i$ with respect to the positive $x$-axis, where $\theta_i$ is uniformly distributed in $[0, 2\pi)$, $i \in \{1, 2, 3\}$. Let $v^{*}$ and $\theta^{*}$ denote the speed and direction of the equivalent serving UAV, respectively. By using elementary geometric theory, we obtain 
\begin{equation} \label{Eq-9}
	v^{*} = \left({\left(\sum_{i=1}^{3}v_i \cos{\theta_i}\right)^2+\left(\sum_{i=1}^{3}v_i \sin{\theta_i}\right)^2}\right)^{1/2},
\end{equation}
and  
\begin{equation} \label{Eq-10}
	\theta^{*} = \arctan\left(\frac{\sum_{i=1}^{3}v_i \sin{\theta_i}}{\sum_{i=1}^{3}v_i \cos{\theta_i}}\right).
\end{equation}
Next, the distributions of $v^{*}$ and $\theta^{*}$ are derived and summarized in the following lemma.
\begin{subequations}
\begin{lemma}\label{Lemma_PDF_V_theta} \label{Lemma_PDF_theta}
	The PDF of $v^{*}$ defined by \eqref{Eq-9} is expressed as
	\begin{equation} \label{Eq_V*}
	f_{v^{*}}(x) = \frac{2}{\Gamma(a) \, b^a}x^{2a-1} \exp{\left(-\frac{x^2}{b} \right)}, 
	\end{equation} with 
\begin{align}
	a &\triangleq \frac{3\left(\mathbb{E}\left(v_i^2\right)\right)^2}{\mathbb{E}\left(v_i^4\right)+\left(\mathbb{E}\left(v_i^2\right)\right)^2}, \\
	b &\triangleq \frac{\mathbb{E}\left(v_i^4\right)+\left(\mathbb{E}\left(v_i^2\right)\right)^2}{\mathbb{E}\left(v_i^2\right)},
\end{align}
and $\theta^{*}$ defined by \eqref{Eq-10} is uniformly distributed within $[0, 2\pi)$.
\end{lemma}
\end{subequations}

\begin{proof}
	See Appendix~\ref{Proof_PDF_Theta}.
\end{proof}

As a special case, assuming the speed $v_i$ in \eqref{Eq-9} follows Rayleigh distribution with parameter $\sigma$, that is, the average speed is determined as $\bar{v}_i = \sqrt{{\pi}/{2}} \, \sigma$, \eqref{Eq_V*} reduces to
\begin{align}
		f_{v^{*}}(x) = \frac{x}{3\sigma^2} \exp\left(-\frac{x^2}{6\sigma^2}\right),
\end{align}
which is cleary a Rayleigh distribution with parameter $\sqrt{3}\,\sigma$. Therefore, we may safely arrive at the following corollary.
\begin{corollary}\label{Corollary_PDF_v}
	 When UAVs move according to the DMS mobility model and their speeds are Rayleigh distributed with parameter $\sigma$, the equivalent running speed $v^*$ is also Rayleigh distributed but with parameter $\sqrt{3}\,\sigma$.
\end{corollary}

\begin{figure}[!t]
	\centering
	\includegraphics [width=1.8in, clip]{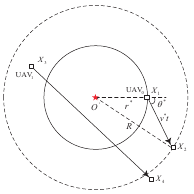}
	\caption{The equivalent UAV (say, $\text{UAV}_0$) moves from $X_1$ with speed ${v^*}$ and angle $\theta^*$ during time $t$ to $X_2$. A handoff occurs if another equivalent UAV (e.g., $\text{UAV}_1$ at $X_3$) is closer than $R$ to $O'$.}
	\label{Fig-6}
\end{figure}

With Lemma~\ref{Lemma_PDF_theta}, we are ready to derive the CoMP handoff probability. As shown in Fig.~\ref{Fig-6}, let $X_1$ be the initial location of the equivalent serving UAV with distance $r^{*}$ from the projected UE location $O'$. Assume that the UAV moves to a new location $X_2$ during time $t$ with speed $v^{*}$ and direction $\theta^{*}$. By recalling the law of cosines, $X_2$ is at a distance 
\begin{equation} \label{Eq-Distance1}
	R = \left({{r^{*}}^2+{v^*}^2t^2+2r^{*} {v^*}t \cos{\theta^*}}\right)^{1/2}
\end{equation}  
from $O'$. Consequently, a handoff occurs if there exits another equivalent UAV (e.g., $\text{UAV}_1$ at $X_3$ in Fig.~\ref{Fig-6}) closer than $R$ to $O'$. Let $c(O', R)$ denote a circle with radius $R$ centered at $O'$. Define $G$ as the event that there is no UAV in $c(O', R)$ and $\bar{H}_1(t)$ as the event that no handoff has occurred until time $t$. We know $\bar{H}_1(t) \subset G$ because of the different speeds of equivalent serving UAVs and the possible event that a UAV can enter and exit the circle $c(O', R)$ during time $t$, e.g., an interfering $\text{UAV}_1$ moving from $X_3$ to $X_4$ in Fig.~\ref{Fig-6}. Consequently, we get $\mathbb{P}\left[\bar {H_1}(t) \right] \leq \mathbb{P}[G]$, which gives a lower bound on the handoff probability, as formalized below.
\begin{theorem}\label{Theorem_DMS_Haoff}
	In the DMS model, the lower bound on the CoMP handoff probability can be computed as 
	\begin{align} \label{Eq_Hanoff_DiffSpeed}
		\mathbb{P}[H_1(t)]  & \geq 1-\int_{0}^{\infty}\int_{0}^{\infty}\int_{-\pi}^{\pi}  \hspace{-0.5em} 2\lambda_0 r^*\exp\left(-2\lambda_0\pi {r^*}^2 \right) f_{v^*}(x) \nonumber\\
		&\quad  \times \exp\left[-\int_{0}^{R} \hspace{-0.5em} 4\pi r \lambda\left(t; r, r^{*}\right) \mathrm{d}r\right] {\mathrm{d}}\theta^{*} {\mathrm{d}}r^{*} {\mathrm{d}} x ,
	\end{align}
	where $\lambda\left(t;r,r^{*}\right)$ is given by \eqref{Lambda_DiffeSpeed} with  $f_{v^*}(x)$ in place of $f_v(x)$. 
\end{theorem}
\begin{proof}
	Given $\{r^{*}, \theta^{*}, v^{*}\}$, the conditional handoff probability can be computed as
	\begin{align}
		\mathbb{P}[H_1(t)|r^{*}, \theta^{*}, v^{*}] 
		& = 1- \mathbb{P}\left[\hat{H}(t)\right] \nonumber \\
		& \geq 1-\mathbb{P}[G] \nonumber \\
		&=1- \mathbb{P}[N(c\left[O', R\right]) =0] \nonumber \\
		&=1-\exp\left[\int_{0}^{R} \hspace{-0.5em} 4\pi r \lambda\left(t;r,r^{*}\right) \mathrm{d} r\right]. \label{Eq-16}
	\end{align} 
 Integrating~\eqref{Eq-16} with respect to $r^{*}$, $\theta^{*}$, and $v^{*}$ yields \eqref{Eq_Hanoff_DiffSpeed}.
\end{proof}

\subsubsection{\underline{Handoff Probability Regarding the SMS Model}}
\label{Handoff_Statical_Scenary}
For comparison purposes, we now calculate the CoMP handoff probability for the SMS model, where all UAVs move at a constant speed ($v_0$) along a straight line in a randomly chosen direction ($\theta_i$), which is uniformly distributed in $[0, 2\pi)$, for all $i \in \{1, 2, 3\}$. Let $v'$ and $\theta'$ denote the speed and direction of the equivalent serving UAV, respectively. By using elementary geometric theory, we obtain 
\begin{equation} \label{Eq-17}
	v'  = v_0\left({\left(\sum_{i=1}^{3} \cos{\theta_i}\right)^2+\left(\sum_{i=1}^{3} \sin{\theta_i}\right)^2}\right)^{1/2},
\end{equation}
and  
\begin{equation} \label{Eq-18}
	\theta' = \arctan\left(\frac{\sum_{i=1}^{3} \sin{\theta_i}}{\sum_{i=1}^{3} \cos{\theta_i}}\right).
\end{equation}

By using a similar approach as in the proof of Lemma~\ref{Lemma_PDF_theta} (cf. Appendix~\ref{Proof_PDF_Theta}), the PDFs of $v'$ in \eqref{Eq-17} and $\theta'$ in \eqref{Eq-18} can be easily derived, as summarized in the following lemma.
\begin{lemma}\label{Lemma_v_theta}
	The PDF of $v'$ defined by \eqref{Eq-17}  is expressed as
	\begin{equation}\label{Eq_V'_SMS}
		f_{v'}(x) = \sqrt{\frac{2}{\pi}}\frac{1}{v_0^3} x^2\exp{\left(-\frac{x^2}{2v_0^2}\right)},~~0\leq x \leq 3v_0
	\end{equation} 
and $\theta'$ defined by \eqref{Eq-18} is uniformly distributed in $[0, 2\pi)$.
\end{lemma}

By recalling the displacement theorem for PPP \cite[Theorem 2.33]{Haenggi12}, the cooperative network under study remains a homogeneous PPP with density $\lambda_0$. Therefore, the intensity of interfering UAVs is given by 
\begin{equation}\label{Eq_Interference_Lambda}
	\lambda(r_i, r_3) =
	\left\{\begin{array}{rl}
		\lambda_0, & r_i > r_3;\\
		0, & r_i \leq r_3.
	\end{array} \right.
\end{equation}

With Lemma~\ref{Lemma_v_theta} and \eqref{Eq_Interference_Lambda}, using a similar equivalent handoff model to the DMS model, we can readily derive the CoMP handoff probability of the SMS model. Since a handoff event is height-free (note that all UVAs under study are supposed to fly at the same height) and depends solely on the characteristics of the UAV point process at a specific time $t$, the CoMP handover probability regarding the SMS model is the same as that of a dual equivalent static UAV model, in which UAVs follow a homogeneous PPP and the typical UE moves at a random velocity characterized by $v'$ and $\theta'$. 

\begin{figure}[!t]
	\centering
	\includegraphics [width=1.8in, clip]{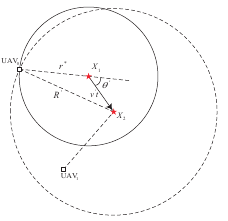}
	\caption{The typical projected UE locating at $X_{1}$ with a distance $r^{*}$ from the equivalent serving UAV (say, $\text{UAV}_0$), moves a distance ${v'}t$ with angle $\theta'$ to $X_2$. A handoff occurs if there is another equivalent UAV (say, $\text{UAV}_1$) closer than $R^{\prime}$ to the UE at $X_2$.}
	\label{Fig-7}
\end{figure}

As illustrated in Fig.~\ref{Fig-7}, the typical projected UE, originating from $X_{1}$ at a distance $r^{*}$ away the equivalent serving $\rm{UAV}_0$, moves at an equivalent speed $v'$ to $X_2$ with a distance ${v'}t$. By using the law of cosines, $X_2$ is at a distance
\begin{equation} \label{Eq-Distance2}
	R' = \left({r^{*}}^2+{v'}^2t^2+2r^{*} {v'}t \cos{\theta'} \right)^{1/2}
\end{equation}
from the UE. As a result, a handoff occurs if there is another equivalent UAV (e.g., $\rm{UAV}_1$ in Fig.~\ref{Fig-7}) closer than $R'$ to the projected UE at the new location $X_2$. Since the typical projected UE moves a distance ${v'}t$, the conditional handoff probability can be derived, as summarized in the following lemma.

\begin{subequations}
	\begin{lemma}\label{Lemma_1}
		Given $\{r^{*}, v', \theta'\}$, a lower bound on the conditional handoff probability can be expressed as 
		\begin{align} \label{Equation_Conditional_Handoff}
			\lefteqn{\mathbb{P}[H_2(t)|r^{*}, v', \theta']} \nonumber \\
			& \geq 1-\exp\left(2\lambda_0{r^{*}}^2 g(\phi_1) - 2\lambda_0 R'^2\left(\pi - g(\phi_2)\right)\right),
		\end{align}
		where
		\begin{align}
			g(x) &\triangleq x-  \sin(2x)/2, \label{Eq-21b} \\
			\phi_1 &\triangleq \arccos\left(\left({r^{*}}^2 + {v'}^2t^2 - {R'^2}\right)/\left(2r^{*}v't\right)\right),  \label{Eq-21c} \\
			\phi_2 &\triangleq \arccos\left(\left(R'^2+ {v'}^2t^2 -{r^{*}}^2\right)/\left(2R' v't\right) \right). \label{Eq-21d} 
		\end{align}
	\end{lemma}
	\begin{proof}
		See Appendix~\ref{Proof_Lemma_1}.
	\end{proof}
\end{subequations}

In light of Lemma~\ref{Lemma_1}, by taking the average with respect to $r^{*}$, $v'$, and $\theta'$, the lower bound on the CoMP handoff probability can be expressed as
\begin{align}\label{Eq_HandoffPr_Lower} 
	\lefteqn{ \mathbb{P}[H_2(t)] } \nonumber\\
	& \geq 1- \frac{1}{2\pi} \int_{0}^{3v_0} \int_{0}^{\infty}\int_{-\pi}^{\pi}  4\lambda_0\pi r^{*}f_{v'}(x)  \exp\left(-2\lambda_0  \right. \nonumber\\ 
	&\quad \times \left.\left[{r^{*}}^2 \left(\pi - g(\phi_1)\right) +R'^2\left(\pi-g(\phi_2 )\right)\right]  \right)     \mathrm{d}\theta' \mathrm{d}r^{*} \mathrm{d}x.
\end{align}
Then, by using a similar approach as that in \cite{8673556}, \eqref{Eq_HandoffPr_Lower} can be computed as (for more details, please refer to Appendix~\ref{Proof_Eq_Handoff_1})
\begin{subequations} \label{Eq-23}
	\begin{equation} \label{Equation_Handoff_1}
		\mathbb{P}[H_2(t)] \geq 1-4\lambda_0 (J_1+J_2+J_3),
	\end{equation}
	where 
	\begin{align}
		J_1 &\triangleq  \int_{0}^{3v_0} \int_{0}^{\infty} \int_{0}^{{\pi}/{2}}\hspace{-5pt} q(u_1, r^{*}, \theta', v')f_{v'}(x)  {\rm{d}}\theta' {\rm{d}}r^{*}  {\rm{d}}x, \\
		J_2 &\triangleq \int_{0}^{3v_0} \int_{0}^{\delta}  \int_{{\pi}/{2}}^{\pi} q(u_2, r^{*}, \theta', v')f_{v'}(x)  \, {\rm{d}}\theta' {\rm{d}}r^{*}  {\rm{d}}x, \\
		J_3 &\triangleq  \int_{0}^{3v_0} \int_{\delta}^{\infty} \int_{{\pi}/{2}}^{\pi} q(u_1, r^{*}, \theta', v')f_{v'}(x)  \, {\rm{d}}\theta' {\rm{d}}r^{*}  {\rm{d}}x,
	\end{align}
	with
	\begin{align}
		{q(x, r^{*}, \theta', v')} &\triangleq r^{*} \exp(-2\lambda_0(R'^2 x^2+{r^*}^2\theta' + r^{*}v't \sin{\theta'})), \\
		\delta &\triangleq v't\cos(\pi-\theta'),  \\
		u_1 &\triangleq \pi-\theta'+\arcsin({v't\sin{\theta'}}/{R'}), \\
		u_2 &\triangleq 2\pi-\theta'-\arcsin({v't\sin{\theta'}}/{R'}).
	\end{align}
\end{subequations}

\subsubsection{\underline{Coverage Probability with Handoffs}}
According to the strategy that each UE chooses its CoMP set (cf. Fig.~\ref{Fig_2}), the distances between each UE and its three serving UAVs can be approximated to the first three nearest ones $\{d_1, d_2, d_3\}$ \cite{9893878}. Then, the SIR computed by \eqref{Eq_RxSINR} reduces to
\begin{equation} \label{Eq_RxSINR-CaseI} 
	\Upsilon(0) =\frac{S(0)}{I(0)}\approx \frac{\left| \sum\limits_{i=1}^{3}d^{-\frac{\alpha}{2}}_i \|{g}_i\| \right|^2}{\sum\limits_{j =4}^{\infty} d^{-\alpha}_{j, \, 0} h_j }.
\end{equation}

By using the Palm distribution theory, the joint distribution of the distance parameters $r_1$, $r_2$, and $r_3$ is obtained in \cite{Moltchanov2012Distance}, reproduced as
\begin{equation} \label{Eq_JointPDF_TypeI}
	f_{r_{1}, r_{2}, r_{3}}(x, y, z) = (2\lambda \pi)^3 x y z \exp\left(-\lambda \pi z^2\right).
\end{equation}
With the intensity shown in \eqref{Eq_Interference_Lambda} and the joint PDF of $\{r_1, r_2, r_3\}$ given by \eqref{Eq_JointPDF_TypeI}, the coverage probability of the typical UE can be computed, as formalized below.
\begin{subequations}
\begin{lemma} \label{Lemma_2}
	Given a prescribed outage threshold $\gamma$, an upper bound on the coverage probability of the typical UE is
	\begin{eqnarray} \label{Eq_Theorem_2}
		\mathbb{P}_{\rm C}
		\leq \hspace{-12pt}\int\limits_{0<{r_1} \leq r_2\leq r_3<\infty} \hspace{-20pt}\left \| \exp(\bm{E}(d_i))  \right \|_1 f_{r_{1}, r_{2}, r_{3}}(x, y, z)  {\rm d} {x}{\rm d}{y}{\rm d}{z}, 
	\end{eqnarray}
	where $\bm{E}(d_i)$ is a $\varepsilon \times \varepsilon$ lower triangular Toeplitz matrix, expressed as
	\begin{equation}\label{Eq_ToplitzQMatrix}
		\bm{E}=\left[ \begin{matrix}
			e_0\\e_1&e_0\\e_2&e_1&e_0 \\ \vdots & \vdots& \vdots &\ddots \\ e_{\varepsilon-1}&\cdots &e_2&e_1 &e_0
		\end{matrix}\right],
	\end{equation} 
	with 
	\begin{align} \label{Eq_varpi}
		\varepsilon = {\rm round} \left( \mathbb{E}_{\{r_1, r_2, r_3\}} \left[\frac{n_1\left(\sum_{i=1}^{3}d_i^{-\alpha}\right)^2}{\sum_{i=1}^{3}\left(d_i^{-\alpha} \right)^2} \right]\right),
	\end{align}
	 and the entry $e_k$, $k = 0,1,\cdots, \varepsilon-1$, given by
	\begin{align} \label{Eq_q_n}
		e_k &= \lambda \pi {d_3^2} \delta(k) -\frac{2\Gamma(k+n_2)\lambda \pi d_3^{2-k\alpha} \left({\beta_2\gamma}\right)^k}{(2-k\alpha)\Gamma(n_2)k!  \left({3\beta'}\right)^{k}}  \nonumber \\
		& \quad \times \pFq{2}{1}{k+n_2, k-\frac{2}{\alpha}}{k+1-\frac{2}{\alpha}}{-\frac{\beta_2  \gamma d_3^{-\alpha}}{3 \beta'}}.
	\end{align}
\end{lemma}
\end{subequations}
\begin{IEEEproof}
	See Appendix~\ref{Proof_Lemma_2}.
\end{IEEEproof}

By substituting \eqref{Eq_Hanoff_DiffSpeed} regarding the DMS model (or \eqref{Eq_HandoffPr_Lower} regarding the SMS model) and \eqref{Eq_Theorem_2} into \eqref{Eq_Coverage_Handoff_form2}, an upper bound on the coverage probability with handoffs can be derived, as formalized below. 
\begin{theorem}\label{Theorem_2}
	In light of the handoff probability given by \eqref{Eq_Hanoff_DiffSpeed} regarding the DMS model or \eqref{Eq_HandoffPr_Lower} regarding the SMS model, and the coverage probability given by \eqref{Eq_Theorem_2}, the typical UE's coverage probability with handoffs is upper bounded by 
	\begin{equation}\label{Theorem_Coverage_Handoff}
		\mathcal{P}(t) \leq \left((1-\zeta)+ \zeta \left(1-\mathbb{P}{[H_i(t)]} \right)  \right) \mathbb{P}_{\rm C}, \quad i = 1, 2.
	\end{equation}
\end{theorem}

Finally, we discuss the special case when the speed of the UAV CoMP set is constant. In this case, there is no probable event in which a UAV can enter and exit the handoff area before time $t$, which differs from the DMS scenario. Thus, the lower bound \eqref{Eq_Hanoff_DiffSpeed} degenerates to the exact result. Specifically, substituting $f_{v^*}(x)=\delta(\bar{v}-x)$, where $\delta(\cdot)$ denotes the unit impulse function, into \eqref{Eq_Hanoff_DiffSpeed}  and after some mathematical manipulations, we have 
\begin{align} 
	\lefteqn{\mathbb{P}[H_3(t)] } \nonumber \\
	&= 1-\int\limits_{\bar{v}t}^{\infty}\int\limits_{-\pi}^{\pi}  2\lambda_0 r^*\exp\left(-2\lambda_0 (\pi {r^*}^2+\omega) \right) {\mathrm{d}}\theta^{*} {\mathrm{d}}r^{*} \nonumber \\
	&\quad -\int\limits_{0}^{\bar{v}t}\int\limits_{-\pi}^{\pi} \hspace{-0.5em} 2\lambda_0 {r^*} \exp\left(-2\lambda_0 (\pi {r^*}^2+\pi \left(\bar{v}t-r^*\right)^2+\omega)\right) {\mathrm{d}}\theta^{*} {\mathrm{d}}r^{*}, \label{Eq_28}
\end{align}
where $\omega \triangleq \int_{|\bar{v}t-r^*|}^{R} 2 r \arccos\left[({r^*}^2-r^2-\bar{v}^2t^2\right)/\left(2r^*\bar{v}t)\right]{\mathrm d}r$, and
$R \triangleq ({r^*}^2+\bar{v}^2t^2+2r^*\bar{v}t\cos(\theta^{*}))^{1/2}$.
This result is the handoff probability for a typical UE when all UAVs flying with the same speed $\bar{v}$ serve a stationary UE in a non-cooperative scenario. A similar result was derived in Eq.~(24) of \cite{8673556} for single-tier terrestrial cellular networks with $v=\bar{v}$ and $\lambda = 2\lambda_0$, as reproduced below:
\begin{align}
	\lefteqn{\mathbb{P}[H'(t)] } \nonumber \\
	&= 1-4\lambda_0 \left[ \int_{0}^{\infty}  \int_{0}^{\pi/2} r^{*}\exp\left(-2\lambda_0 \left(R^2\mathcal{A}_1+\mathcal{A}_2\right)\right) {\rm d} {\theta^{*}} {\rm d}r^{*}  \right. \nonumber \\
	&\quad \left. +     \int_{0}^{\bar{v}\cos(\pi-\theta)} \int_{\pi/2}^{\pi} r^{*} \exp\left(-2\lambda_0 \left(R^2\mathcal{A}_3+\mathcal{A}_2\right)\right) {\rm d}\theta^{*} {\rm d}r^{*}  \right. \nonumber\\
	&\quad \left. + \int_{\bar{v} \cos(\pi-\theta^{*})}^{\infty} \int_{\pi/2}^{\pi}  r^{*} \exp\left(-2\lambda_0 \left(R^2\mathcal{A}_1+\mathcal{A}_2\right)\right) {\rm d}\theta^{*}   {\rm d}r^{*}  \right], \nonumber
\end{align}with $\mathcal{A}_1 \triangleq \pi -\theta^{*}+\arcsin\left(\bar{v}\sin(\theta^{*})/R\right)$, $\mathcal{A}_2 \triangleq {r^{*}}^2\theta^{*}+r^{*}\bar{v}\sin(\theta^{*})$, and $\mathcal{A}_3 \triangleq 2\pi -\theta^{*}- \arcsin\left(\bar{v}\sin(\theta^{*})/R\right)$.

\begin{figure}[!t]
		\centering
		\includegraphics [width = 0.38\textwidth, clip, keepaspectratio]{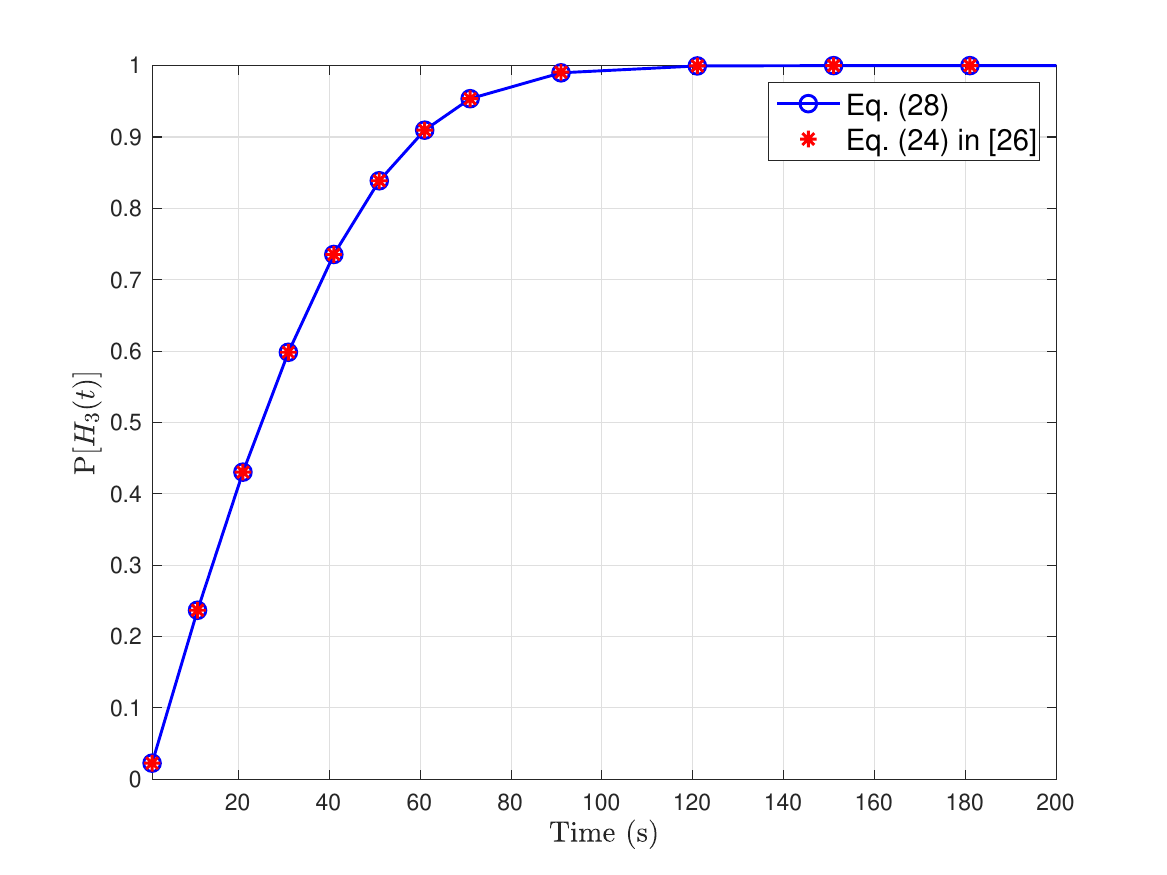}
		\vspace{-10pt}
		\caption{The numerical results of coverage probability computed by \eqref{Eq_28} and \cite[Eq. (24)]{8673556}.}
		\label{Fig-8}
	\end{figure}

The two integral expressions, i.e., \eqref{Eq_28} and [26, Eq. (24)], may appear different and complex at first glance, but computing their numerical results is straightforward using software like Matlab or Mathematica due to the elementary nature of the functions involved in the integrands. Fig.~\ref{Fig-8} shows that their numerical results are identical. This identity is mainly because the handoff event is altitude-free and depends solely on the density of UAVs.

\section{Simulation Results and Discussions}
\label{Section_Simulation}
This section presents and discusses numerical results computed per the previously obtained analytical expressions compared to extensive Monte-Carlo simulation results. We assume that the UAV speed has Rayleigh distribution with parameter $\sigma$ if the DMS movement model is concerned, where the average UAV speed is determined as $\bar{v}_i = \sqrt{{\pi}/{2}} \, \sigma$. Table~\ref{tab1} lists the main simulation parameters.
\renewcommand\arraystretch{1.25}
	\begin{table}[t!]
	\scriptsize
	\vspace{-5pt}
	\caption{Simulation Parameter Setting}
	\vspace{-10pt}
			\begin{center}
			\begin{tabular}{|c|c||c|c|}
				\hline 
				\textbf{Notation} & \textbf{Value} & \textbf{Notation} & \textbf{Value} \\
				\hline
				$\lambda_0$ & $[1, 20]$ \si{UAVs/km^2} & $\zeta$ & $[0, 1]$  \\
				\hline
				$K$& $1 $ & $\sigma$ & $\left[25\sqrt{2/\pi}, 45\sqrt{2/\pi}\right]$ \\
				\hline
				$h$ & 150 \si{m}& $\alpha$ & $[2.4, 3.6]$  \\
				\hline
				$R_s$ & $\geq \sqrt{{18}/{ (\pi \lambda_0)}}$ & $v_0$ & $[25,  45]$ \si{km/h} \\
				\hline
			\end{tabular}
			\label{tab1}
				\end{center}
	\vspace{-5pt}
\end{table}

\subsection{CoMP Handoff Probability}

\begin{figure}[t!] 
	\centering    
	\subfloat[Cooperative handoff probability for the DMS model.] 
	{
		\begin{minipage}[t]{0.5\textwidth}
			\centering         
			\includegraphics[width = 0.925\textwidth]{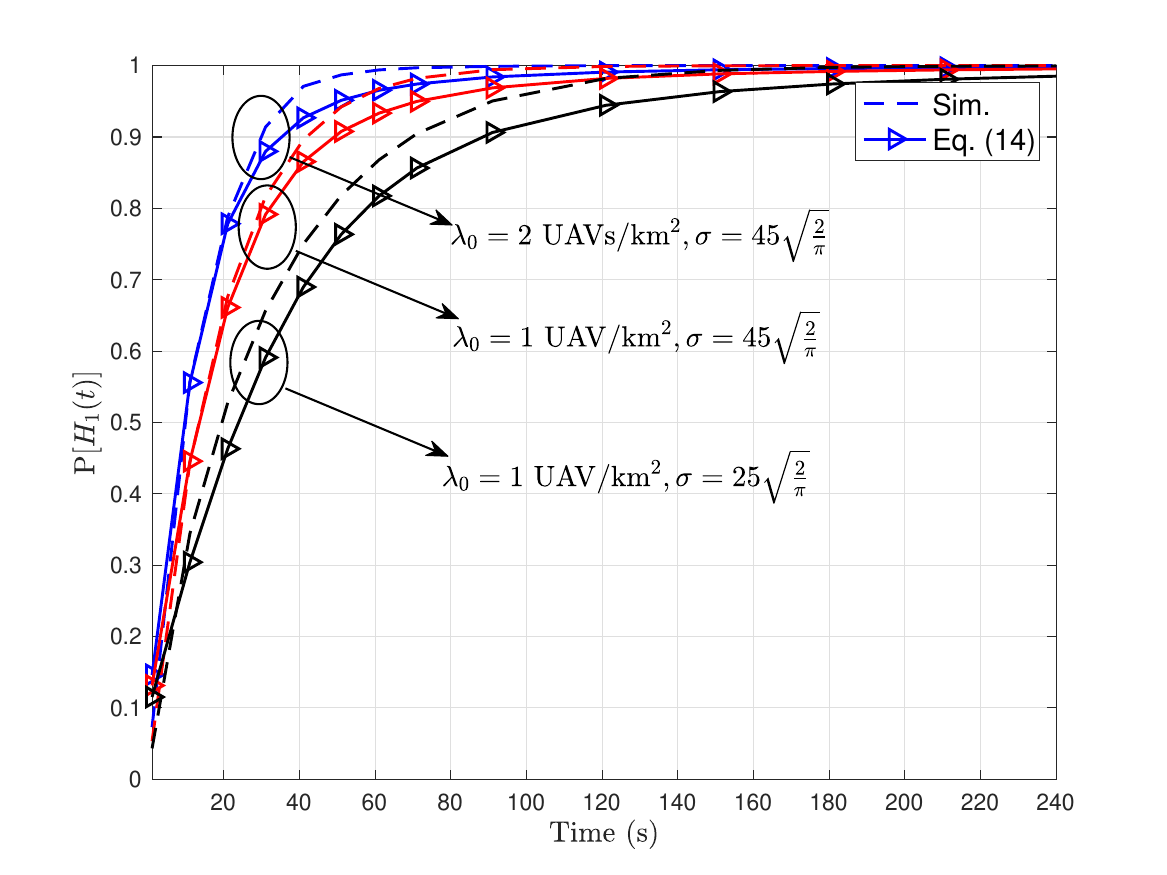}   
			\label{Fig_9_a}
		\end{minipage}
	}

	\subfloat[CoMP handoff probability for the SMS model.] 
	{
		\begin{minipage}[t]{0.5\textwidth}
			\centering      
			\includegraphics[width = 0.925\textwidth]{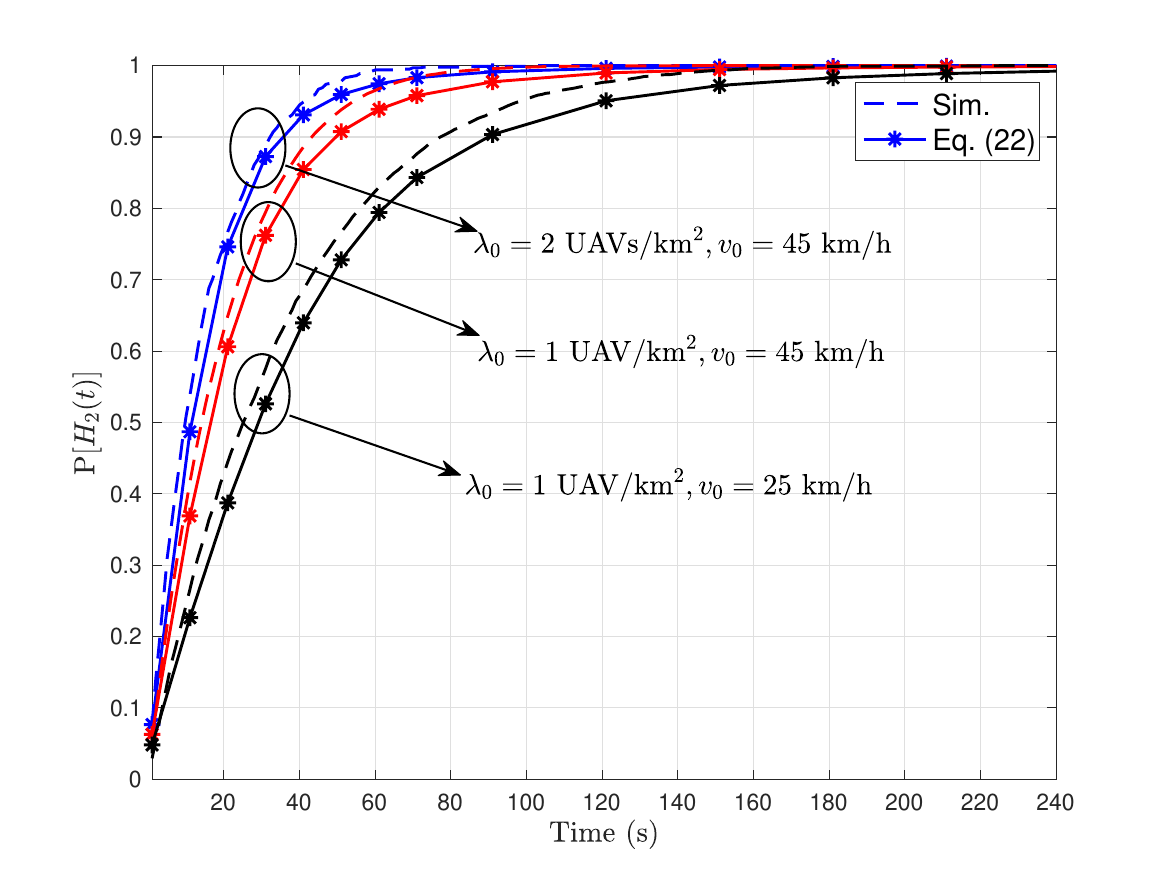}  
			\label{Fig_9_b} 
		\end{minipage}
	}
	\caption{CoMP handoff probabilities of the typical UE under the DMS and SMS movement models.} 
	\label{Fig_9}  
\end{figure}

Fig.~\ref{Fig_9} depicts the CoMP handoff probabilities of the typical UE. In particular, for the DMS model, Fig.~\ref{Fig_9_a} shows the numerical results on the lower bound computed by \eqref{Eq_Hanoff_DiffSpeed}. It is observed that, for a fixed UAV density $\lambda_0 = 1 \si{UAVs/km^2}$, the CoMP handoff probability increases with $\sigma$ (recalling that the average speed of UAVs increases linearly with $\sigma$ in the DMS model). Similarly, given a fixed $\sigma = 45\sqrt{{2}/{\pi}}$, the CoMP handoff probability monotonically increases when the UAV density grows from $\lambda_0 = 1 \ \si{UAVs/km^2}$ to $\lambda_0 = 2 \ \si{UAVs/km^2}$. These observations are due to more frequent handoffs as UAVs fly faster or the cell size shrinks on average, as expected. Similar conclusions can be drawn from Fig.~\ref{Fig_9_b} for the SMS model.

\begin{figure}[t!]
	\centering
	\includegraphics[width=0.5\textwidth, clip]{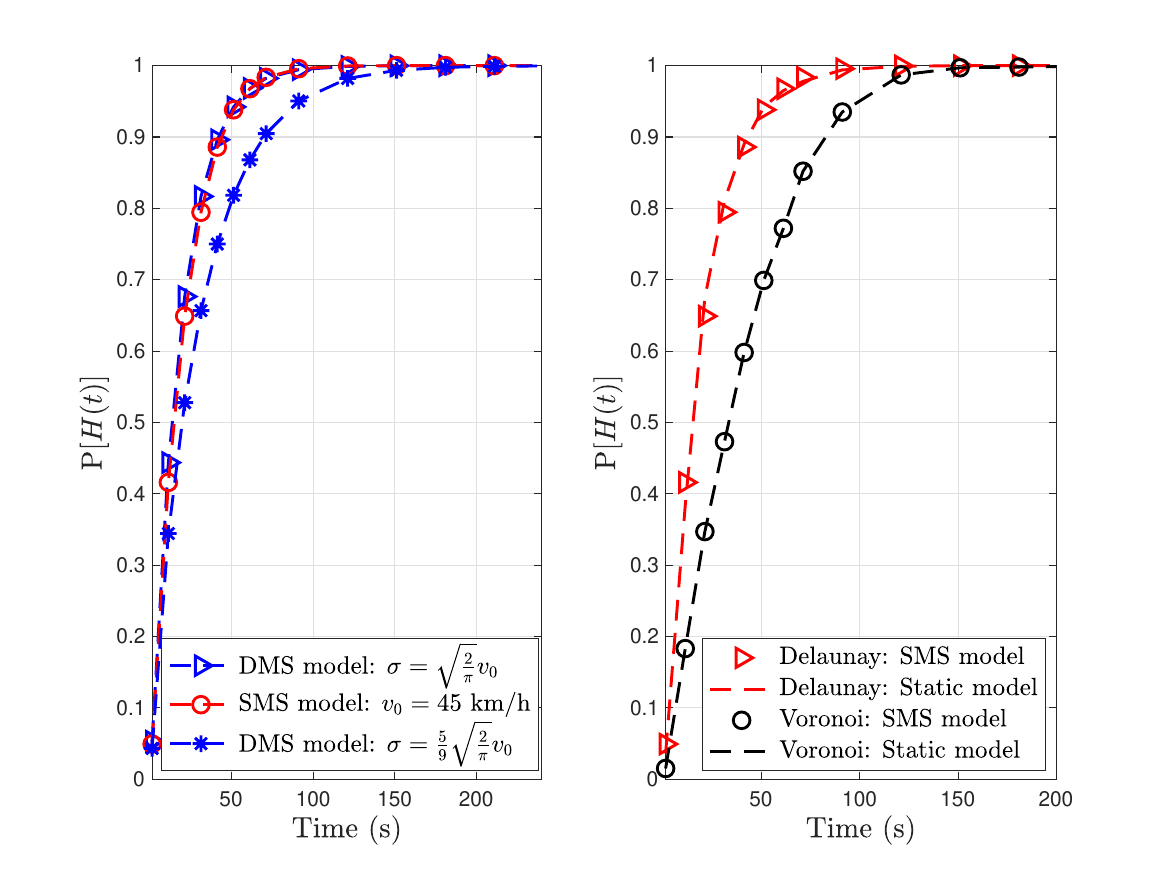}
	\vspace{-10pt}
	\caption{CoMP handoff probability (Left:  DMS and SMS models; Right: SMS and static models under the Delaunay triangulation and Voronoi tessellation)}
	\label{Fig-10}
\end{figure}

The left panel of Fig.~\ref{Fig-10} compares the CoMP handoff probability of the typical UE under the DMS and SMS movement models, where the UAV speed in the DMS model has Rayleigh distribution with parameter $\sigma$ whereas all UAVs move at a constant speed $v_0$ in the SMS model. For comparison purposes, the parameter $\sigma$ in the DMS model is set to $\sqrt{2/\pi} \, v_0$, i.e., the moving speed of UAVs follows a Rayleigh distribution with mean $v_0$. Based on this comparable speed setting, it is interesting to observe that the handoff probability for the DMS model provides a tight upper bound for the SMS model. Meanwhile, when the average speed of UAV is lower than $v_0$, for example, when $\sigma = 5\sqrt{2/\pi} \, v_0/9$, it is noted that the handoff probability of the DMS model is lower than that of the SMS model, as handoffs occur less frequently as UAVs fly slower (in this case, the average UAV speed $\bar{v}_0 = 5{v_0}/9 < v_0$).

The right panel of Fig.~\ref{Fig-10} depicts the CoMP handoff probabilities for the SMS and the equivalent static UAV models under the Delaunay tessellation. In the static UAV scenario, all UAVs follow a homogeneous PPP with density $\lambda_0$, and the typical UE moves along a straight line with an equivalent speed $v'$. We also present the handoff probabilities for both the SMS and static models under Voronoi tessellations for comparison purposes. In these scenarios, the nearest UAV serves each UE for the SMS and static UAV models. In the context of either the Delaunay triangulation or the Voronoi tessellation, it is intriguing to observe that the CoMP handoff probability for the SMS model is identical to that of the equivalent static UAV model. This observation is because the handoff event is independent of UAV height and only depends on the characteristics of the point process of UAVs at time $t$. Meanwhile, the UAVs and BSs in both the SMS and static UAV models adhere to a homogeneous PPP with the same density from the UE perspective at any time $t$. This parallelism leads to the equivalent handoff probabilities in the Delaunay triangulation and Voronoi tessellation. 

\begin{figure}[!t]
	\centering
	\includegraphics[width=0.5\textwidth, clip]{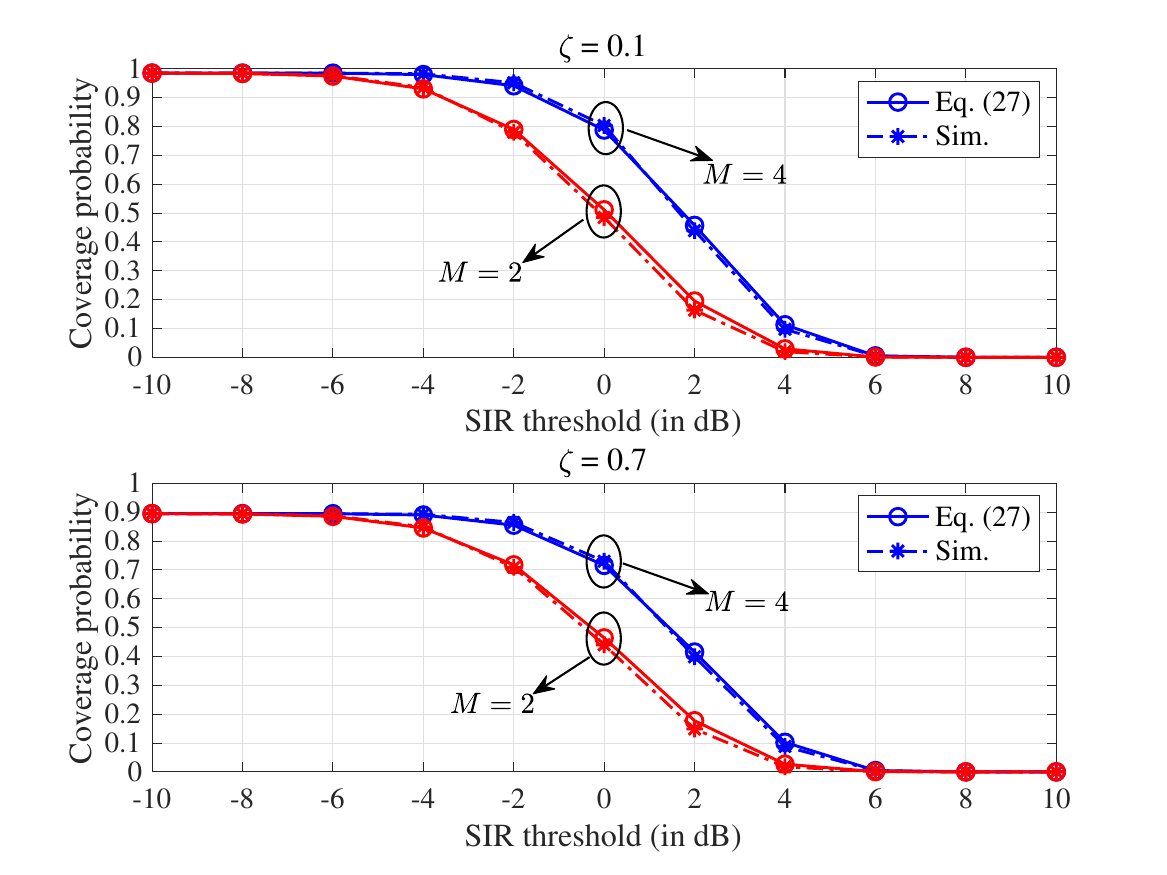}
	\caption{CoMP coverage probability with handoffs versus SIR threshold for the typical UE under the SMS model, with $\alpha = 2.4$ and $\mathbb{P}[H_2(t)] = 0.15$.}
	\label{Fig-11}
\end{figure}

\subsection{Coverage Probability with Handoffs}
 Fig.~\ref{Fig-11} depicts the CoMP coverage probability with handoffs versus the SIR threshold $\gamma$ in dB for the typical UE under the SMS model. In particular, its top panel corresponds to the case of $\zeta = 0.1$ while the bottom panel to the case of $\zeta = 0.7$, where $\zeta$ denotes the probability of connection failure due to handoffs, defined immediately after \eqref{Eq_Coverage_Handoff_form2}. We observe that the CoMP coverage probability with $\zeta = 0.7$ is lower than that with $\zeta = 0.1$ since the system is more sensitive to handoff with a larger $\zeta$. On the other hand, for either case, the coverage probability decreases with the SIR threshold but increases with the number of antennas $M$, as expected. The simulated coverage probability also agrees with the theoretical counterparts, indicating that the proposed approximation method to derive the upper bound is effective.

\begin{figure}[!t]
	\centering    
	\includegraphics[width=0.5\textwidth]{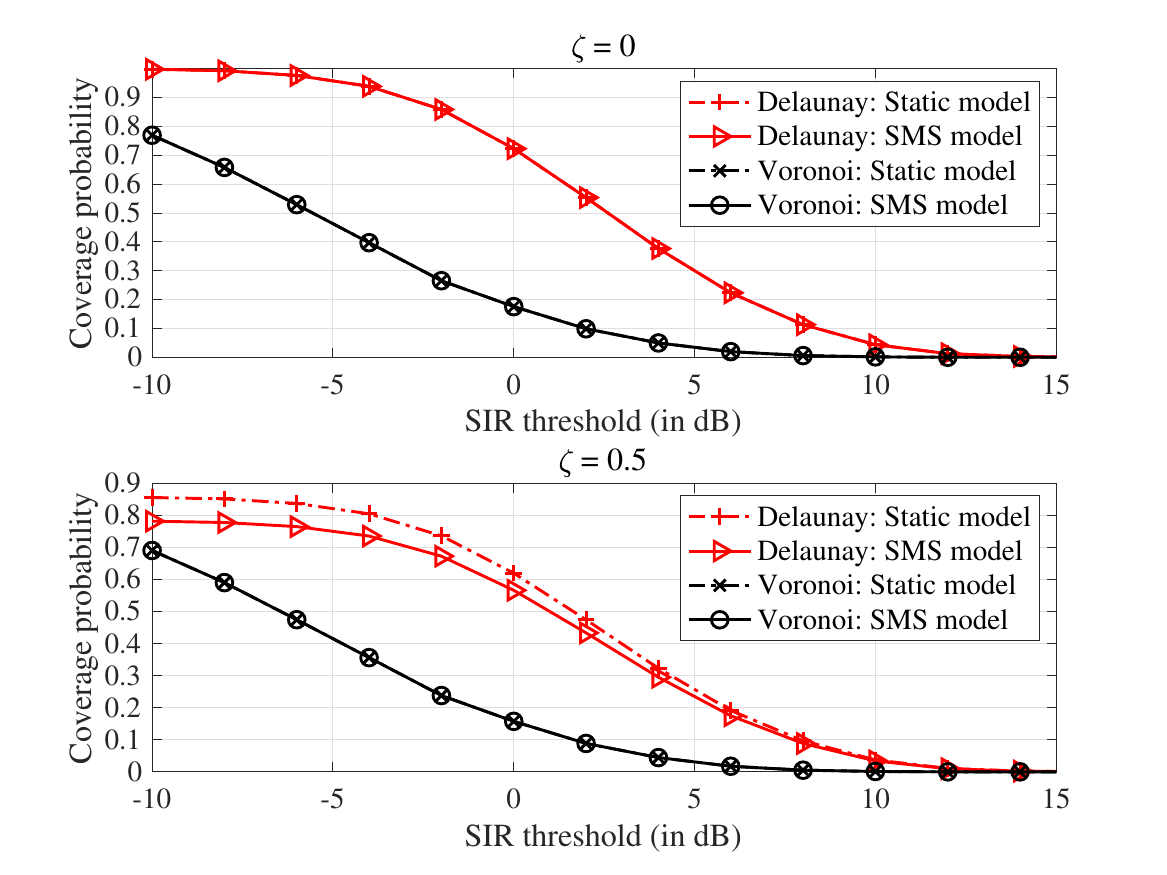}   
	\caption{Coverage probability versus SIR threshold under the proposed Delaunay scheme and the conventional Poisson-Voronoi tessellation scheme, with $\alpha = 2.6$, $t = 3$s, $\lambda_0 = 20~\si{UAVs/km^2}$, and ${v_0}=45$ km/h.} 
	\label{Fig-12}  
\end{figure}

The upper panel of Fig.~\ref{Fig-12} shows the coverage probability of the typical UE versus the SIR threshold $\gamma$ in dB regarding the proposed Delaunay scheme and the conventional Poisson-Voronoi tessellation scheme with the nearest UAV. Both the SMS and static models are considered for both scenarios. The SMS model assumes that all UAVs move at a constant speed $v_0$, while in the static model, all UAVs follow a homogeneous PPP, and the typical UE moves at the same constant speed $v_0$. Given that no handoff is accounted for (i.e., $\zeta = 0$), the upper panel of  Fig.~\ref{Fig-12} shows that the coverage probability of the SMS model is equivalent to that of the static model under the proposed Delaunay scheme and the conventional Poisson-Voronoi scheme. This equivalence is because the spatial distribution of the UAV nodes remains the same at any time in both schemes. 

The lower panel of Fig.~\ref{Fig-12} also shows that, if handoffs are accounted for, the static model has a higher coverage probability than the SMS model with the proposed Delaunay scheme at $\zeta = 0.5$. Although the coverage probabilities without handoffs are equivalent, the handoff probability of the SMS model is higher than the static model, yielding lower coverage probability with handoffs. The higher handoff probability comes from the fact that the CoMP set in the SMS model introduces speed correlation, and the speed $v'$ computed by \eqref{Eq-17} is no longer the constant speed $v_0$. Finally, whether a handoff occurs, the proposed Delaunay scheme significantly outperforms the conventional Voronoi scheme with the nearest UAV regarding the coverage probability. Although the handoff probability of the Delaunay scheme is higher than that of the Voronoi scheme, their gap is effectively bridged by the CoMP gain in the proposed scheme, demonstrating its ability to mitigate the performance degradation caused by handoffs.

\subsection{Extended Performance Evaluation}	
In this subsection, we will enrich the performance evaluation of the proposed model by discussing diverse performance factors. These factors include dynamic power control, varying UAV heights, the general $\eta-\mu$ fading model, spectral efficiency for a typical UE, and a comparison with the conventional hexagonal scheme.

\subsubsection{Effect of Dynamic Power Control}
\begin{figure}[!t]
	\centering
	\includegraphics [width=0.5\textwidth, clip, keepaspectratio]{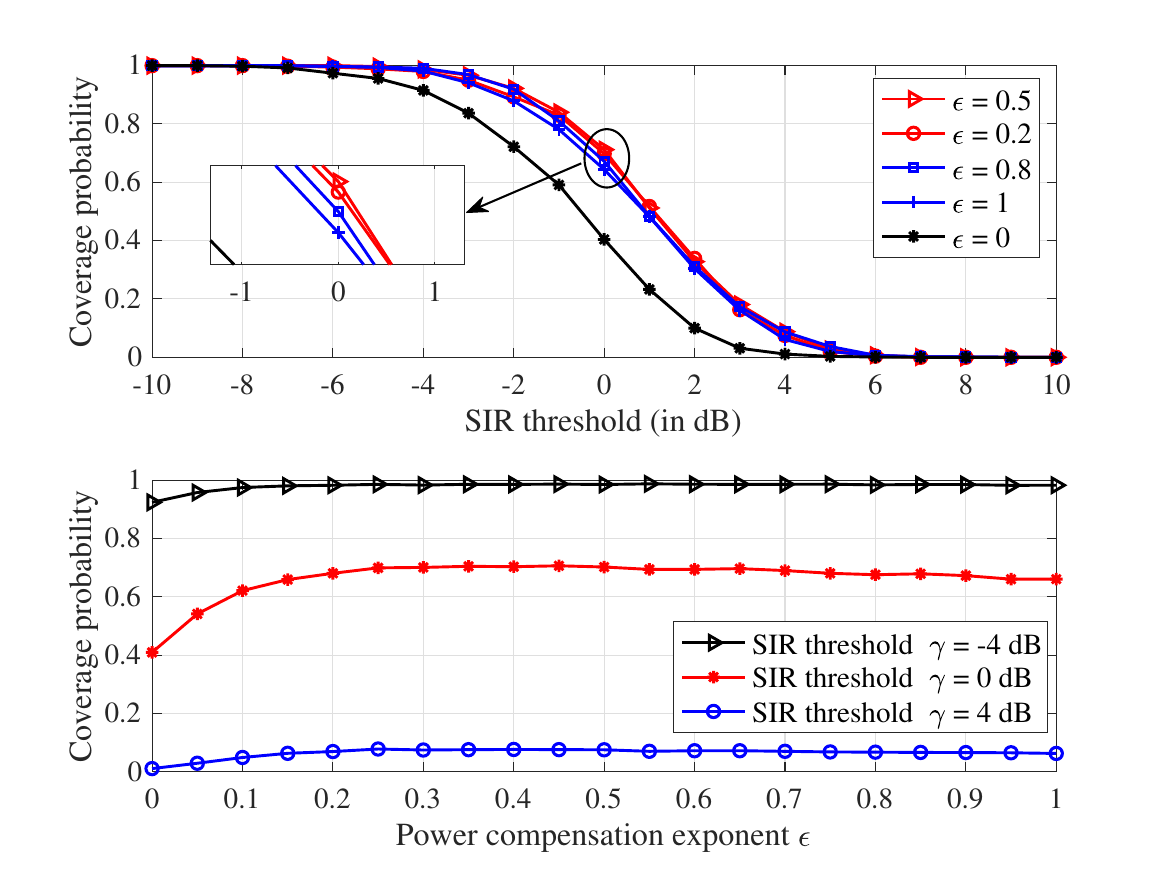}
	\caption{Effect of dynamic power control on the coverage probability.}
	\label{Fig-13}
\end{figure}

It is assumed in Section~\ref{Section_SystemModel} that all UAVs have the same transmit power. However, the proposed Delaunay model can also apply to the case of dynamic power control. Specifically, let $d_i$ denote the distance from UAV $i$ to the typical UE, where $i \in \{\Phi(0) \setminus U(0)\}$. Each UAV $i$ uses the fractional power control strategy, i.e., $P_i = (d_{i}^{\alpha})^{\epsilon}$, where the transmit power per unit distance is normalized and $\epsilon \in [0,1]$ denotes the power compensation exponent. Consequently, the SIR of the typical UE given by \eqref{Eq_RxSINR-CaseI}  becomes
\begin{equation} \label{Eq_SIR_SMS_PowerControl}
	\Upsilon(0) \approx \frac{\left| \sum\limits_{i=1}^{3} \left(d_{i}^{\frac{\alpha}{2}}\right)^{\epsilon} d^{-\frac{\alpha}{2}}_i \|{g}_i\| \right|^2}{\sum\limits_{j =4}^{\infty} (d_{j, \, 0}^{\alpha})^{\epsilon} \, d^{-\alpha}_{j, \, 0} h_j }. 
\end{equation}
With \eqref{Eq_SIR_SMS_PowerControl} and recalling the definition of coverage probability, we can obtain the coverage probability with dynamic power control, plotted in the upper panel of Fig.~\ref{Fig-13}. It is observed that the coverage probability can be improved by dynamic power control, compared to the case with fixed transmit power (i.e., $\epsilon = 0$ in the figure). However, different values of $\epsilon$ have little effect on the amount of performance improvement. The main reason is as follows: The fractional power control strategy aims to eliminate path loss and has similar effects on the desired signal power and interference power. Therefore, it has little effect on the received SIR and the coverage probability. For further illustration, the lower panel of Fig.~\ref{Fig-13} plots the coverage probability versus different power compensation exponent $\epsilon$. It is observed that the coverage probability remains almost fixed when $\epsilon \ge 0.2$ as the path-loss effects of desired signals and interferences are almost canceled.

	\begin{figure}
		\centering
		\includegraphics [width = 0.5\textwidth, clip, keepaspectratio]{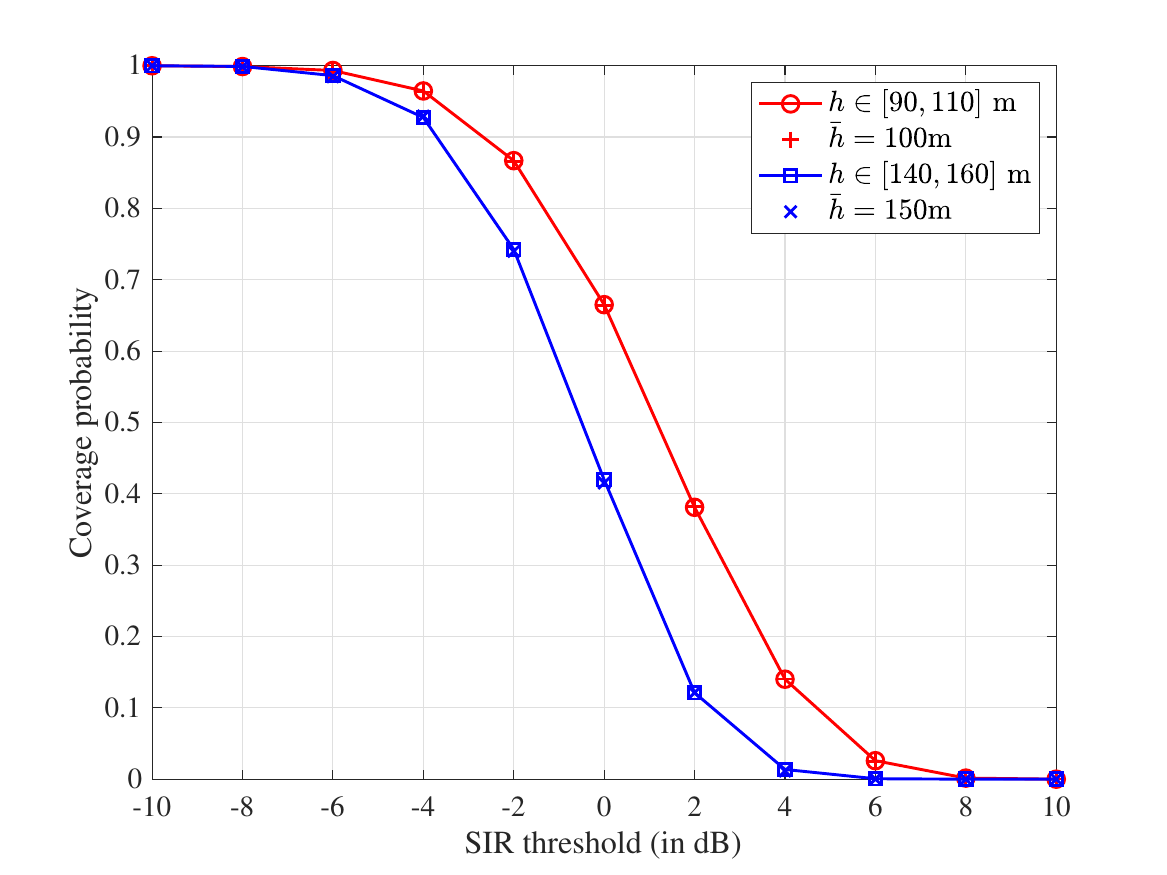}
		\caption{Coverage probability versus SIR threshold for a typical UE with varying UAV heights, with $M = 2$, $\zeta =0$, $\lambda_0 = 20~\si{UAVs/km^2}$, and $\alpha = 2.4$.}
		\label{Fig-14}
	\end{figure}

\subsubsection{Effect of Varying UAV Heights} 
	In Section~\ref{Section_SystemModel}, it is assumed that UAVs hover at the same height. Footnote~\ref{Footnote-1} explains that UAV height difference has little effect on the system performance because it is negligible compared to the distance between UAVs and terrestrial UEs. For further illustration, Fig.~\ref{Fig-14} shows the coverage probability versus SIR under the instantaneous height, $h$, and the average, $\bar{h}$. It is observed that they agree with each other for either the case of $h \in [90,110]$~\si{m} with $\bar{h} =100$~\si{m} or the case of $h \in [140, 160]$~\si{m} with $\bar{h} = 150$~\si{m}. Therefore, the proposed handoff management strategy and associated performance analysis also apply to 3D mobile UAVs.

	\begin{figure}
		\centering
		\includegraphics [width = 0.5\textwidth, clip, keepaspectratio]{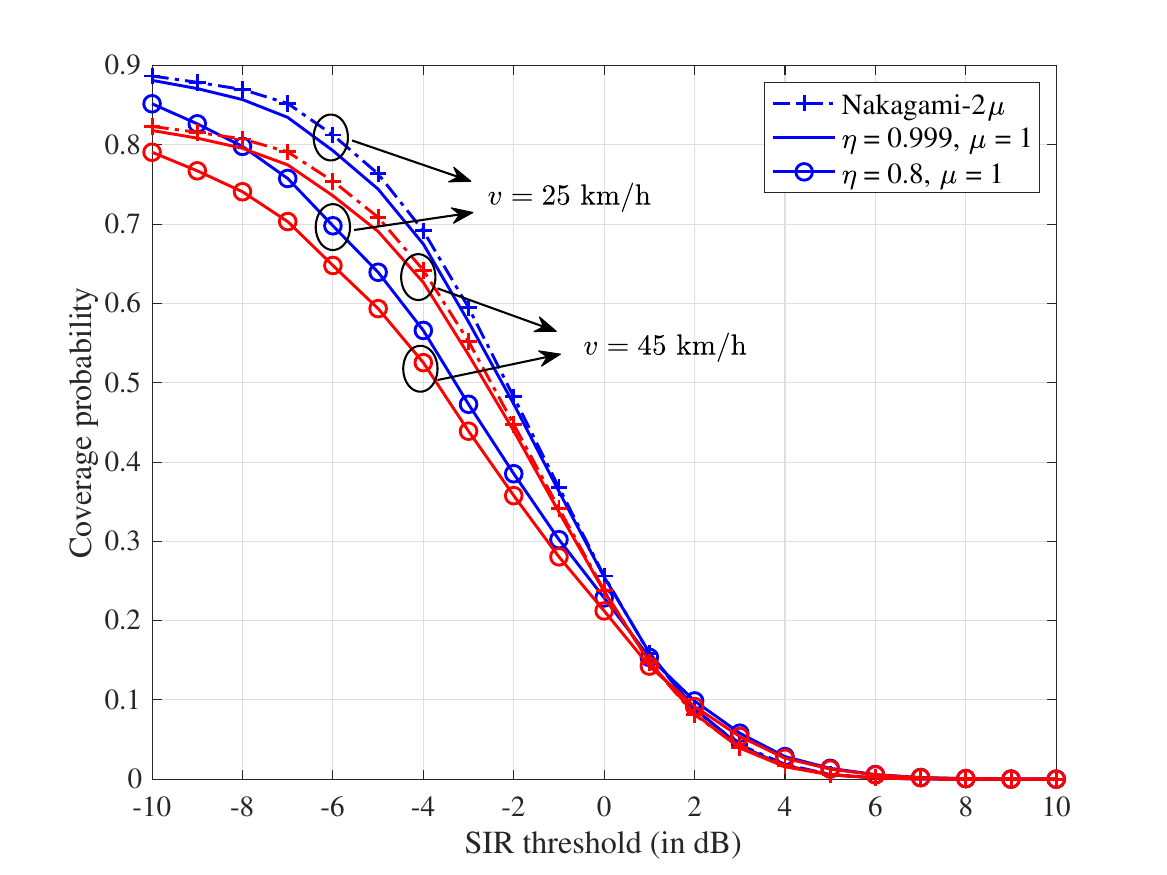}
		 \caption{Coverage probability for a typical UE under the general $\eta-\mu$ fading model, with $\zeta = 0.4$, $t = 3$~\si{s}, $\lambda_0 = 20$~\si{UAVs/km^2}, and $\alpha = 2.4$.}		\label{Fig-15}
	\end{figure}

\begin{figure}
	\centering
	\includegraphics [width = 0.5\textwidth, clip, keepaspectratio]{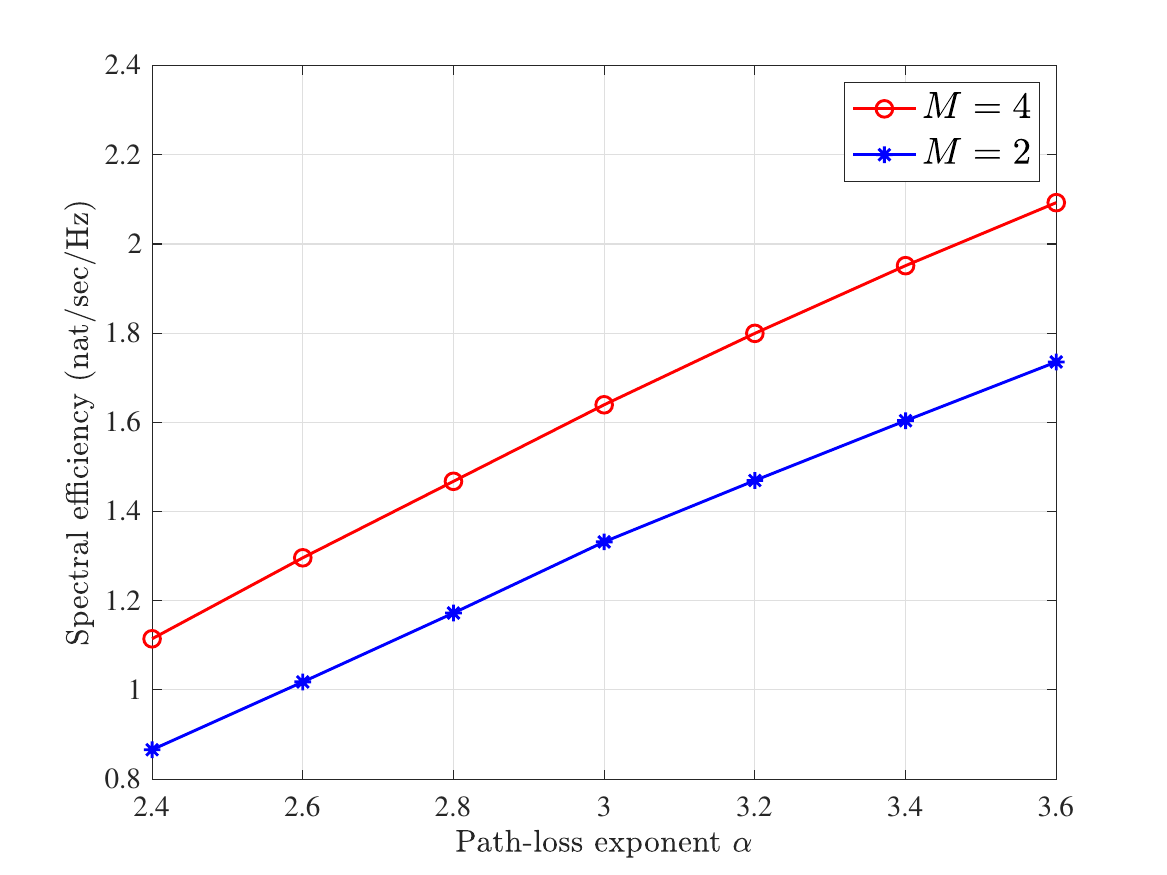}
	\caption{Spectral efficiency of a typical UE versus the PLE $\alpha$, with $\zeta =0$, $\lambda_0 = 20$~\si{UAVs/km^2}, and $h= 150$~\si{m}.}
	\label{Fig-16}
\end{figure}

	\subsubsection{Effect of the General $\eta-\mu$ Fading Model} \label{Subsection-GeneralFading}
	A more general $\eta-\mu$ fading model \cite{887126} is considered in the simulation experiments to characterize the various fading environments. As shown in Fig.~\ref{Fig-15}, on the one hand, for a given $\mu$, the coverage probability with handoffs increases with $\eta$. When $\eta$ approaches $1$ and $m = 2\mu$, the performance in the $\eta-\mu$ fading model is almost the same as that in the Nakagami-$m$ model. In other words, the $\eta-\mu$ distribution degenerates into the Nakagami-$m$ distribution, as disclosed in \cite{887126}. On the other hand, with $\mu$ and $\eta$ fixed, the coverage probability with handoffs decreases as the UAV moving speed increases. The reason is mainly that the increased mobility leads to more frequent handoffs and thus reduces coverage performance.

\subsubsection{Spectral Efficiency for a Typical UE} 
	Fig.~\ref{Fig-16} plots the spectral efficiency of a typical UE versus the PLE $\alpha$. It is seen that the spectral efficiency increases with $\alpha$ because the suffered interference decreases faster than the received signal, though the received signal also decreases with $\alpha$. A similar phenomenon was observed in Fig.~4 of the pioneering work \cite{6042301}. On the other hand, for a fixed $\alpha$, the spectral efficiency increases with more transmitting antennas, as expected.

	 \subsubsection{Comparision with Conventional Hexagonal Models} 	 	 
\begin{figure}[!t]
		\centering
		\includegraphics [width = 0.5\textwidth, clip, keepaspectratio]{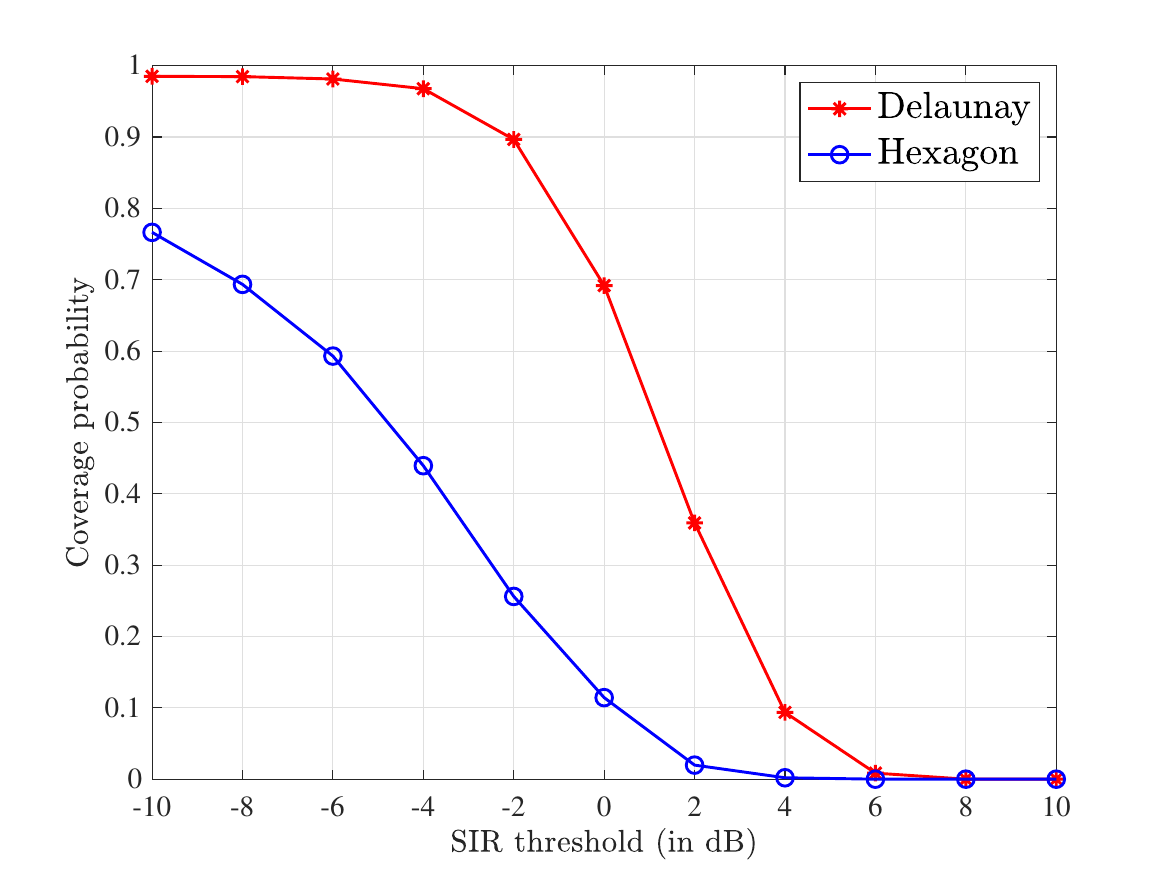}
		\caption{Coverage probability versus the SIR threshold for a typical UE under the proposed Delaunay scheme and the conventional hexagonal scheme, with $\lambda_0 = 20~\si{UAVs/km^2}$, $M = 2$, $\alpha = 2.4$, and $h= 150$ m.}
		\label{Fig-17}
	\end{figure}

	 A conventional CoMP model based on hexagonal cells is considered for comparison purposes \cite{9644611}. In the simulation experiments, the average number of UAVs in each hexagonal cell is assumed to be three for a fair comparison. Given the same parameter settings, the simulation results of the coverage probability with handoffs are shown in Fig.~\ref{Fig-17}, which illustrates that the coverage performance of UEs under our proposed scheme is superior. This is because the distances from the serving UAVs within the conventional hexagonal cell to the typical UE are not the nearest, and its suffering interference is more serious.
	 
\section{Conclusions}
\label{Section_Conc}
This paper developed a coordinated multi-point (CoMP) transmission scheme for air-to-ground (A2G) communications using the principle of stochastic geometry regarding Poisson-Delaunay triangulation. In particular, each terrestrial UE is jointly served by three UAVs forming a triangular cell, thus improving the quality of service compared to the service by a single UAV. In addition, a dynamic subdivision CoMP set search scheme was designed to save searching costs using a recursive divide-and-conquer algorithm. Compared to the conventional Voronoi scheme with the nearest-neighbor association criterion, although the proposed CoMP scheme has a more frequent handoff probability caused by the smaller Delaunay triangular cells than the dual Voronoi cells, it attains a higher coverage probability with handoffs, thanks to the CoMP gain. Consequently, the proposed CoMP transmission scheme is promising for A2G communications to improve network coverage where UAV handoffs occur frequently.

\appendices

\section{Proof of Lemma~\ref{Lemma_PDF_V_theta}}
\label{Proof_PDF_Theta}
Define $v^{*}\triangleq \sqrt{Z} $, where  
\begin{align}
	Z &\triangleq\left(\sum_{i=1}^{3}v_i \cos{\theta_i}\right)^2+\left(\sum_{i=1}^{3}v_i \sin{\theta_i}\right)^2 \nonumber\\
	   &=\sum_{i=1}^{3}v_i^2+\sum_{i=1}^{3}v_i v_j \cos\left(\theta_i-\theta_j\right).
\end{align}

Assuming the speed $v_i$ are i.i.d. with mean $\mathbb{E}(v_i)$ and variance $\text{Var}(v_i)$, the distributions of $v^{*}$ is then derived as follows.

Based on the causal form of the central limit theorem, it is known that the distribution of $z$ can be approximated by the Gamma distribution given by 
\begin{equation}\label{Eq_Z}
	f_Z(z) = \frac{z^{a-1}}{\Gamma(a) \, b^a} \exp\left(-\frac{z}{b}\right)
\end{equation} with the parameters determined by 
\begin{equation}
	a= \frac{\mathbb{E}^2(Z)}{\rm{Var}(Z)}, \,\,\, b = \frac{\rm{Var}(Z)}{\mathbb{E}(Z)},
\end{equation}where
\begin{align}
	\mathbb{E}(Z) = 3\,\mathbb{E}\left(v_i^2\right) 
\end{align}
  and 
 \begin{align}
 	\text{Var}(Z) &= 3 \text{Var}(v_1^2) +6\text{Var}\left(v_1v_2\cos\left(\theta_1-\theta_2\right)\right),\nonumber\\
 	&= 3 \text{Var}(v_1^2) +6 \mathbb{E}\left(v_1^2\right)\mathbb{E}\left(v_2^2\right) \nonumber\\
 	&= 3\mathbb{E}\left(v_i^4\right)+3\left(\mathbb{E}\left(v_i^2 \right) \right)^2.
 \end{align}

By \eqref{Eq_Z}, the CDF of $v^{*}$ can be expressed as 
\begin{align}
	F_{v^{*}}(x) &\triangleq {\rm{Pr}}  \left\{v^{*}\leq x \right\}    =   {\rm{Pr}}\{\sqrt{Z} \leq x \} \nonumber\\
	&= {\rm{Pr}} \left\{Z\leq {{x}^2}\right\} = F_{Z}\left( {x}^2 \right).
\end{align}
Then, taking the derivative of $F_{v^{*}}(x)$ with respect to $x$ yields
\begin{align}
	f_{v^{*}}(x) = 2 x f_Z\left( x^2\right).
\end{align}
After simplifying, we get the desired \eqref{Eq_V*}.

To derive the distribution function of $\theta^{*}$, by using a similar approach as \cite[Appendix C]{9078878}, we introduce two auxiliary random variables $\psi_i = \theta_i$, $i = 1, 2$, and find the joint PDF of three random variables $\theta^{*}$ and $\psi_i$, $i = 1, 2$. We then integrate these auxiliary random variables to find the PDF of $\theta^{*}$. We start by solving the system of three equations (one equation for the definition of $\theta^{*}$ and two equations introduced by the auxiliary random variables) to derive $\theta_i$'s in terms of $\theta^{*}$ and $\psi_i$'s. Finally, we obtain two solution sets: 
	\begin{align} \label{Eq_PDF_theta_1}
	(1) \left\{\begin{array}{lr}
		\theta_i = \psi_i, & i = 1, 2; \\ 
		\theta_3 = \arctan\left(\frac{-A\cos(\theta^*)+A'\sin(\theta^*)}{A\sin(\theta^*)+A'\cos(\theta^*)}\right) \triangleq \psi_{3a}, & \text{otherwise},
	\end{array} \right. 
\end{align}

\begin{align} \label{Eq_PDF_theta_2}
	(2) \left\{\begin{array}{lr}
		\theta_i = \psi_i, &  i = 1, 2; \\ 
		\theta_3 = \arctan\left(\frac{-A\cos(\theta^*)-A'\sin(\theta^*)}{A\sin(\theta^*)-A'\cos(\theta^*)}\right)  \triangleq \psi_{3b}, & \text{otherwise},
	\end{array} \right. 
\end{align}
where $A \triangleq \sum_{i=1}^{2}v_i \sin\left(\psi_i-\theta^*\right)$ and $A' \triangleq \sqrt{v_3^2-A^2}$. Computing the  determinant of the Jacobian matrix $J$, i.e., $|J| = \left|\frac{\partial \theta_3}{\partial \theta^*}\right|$ for both solution sets, we get
\begin{equation}
	|J| = \frac{\sum_{i=1}^{3}\sum_{j=1}^{3} v_i^2\cos(\theta_i-\theta_j)}{\sum_{i=1}^{3} v_i^2\cos(\theta_i-\theta_3)}.
\end{equation}
After performing some algebraic manipulations, we attain
\begin{equation}
	|J|\bigg|_{\theta_3=\psi_{3a}}+|J|\bigg|_{\theta_3=\psi_{3b}}=2.
\end{equation} As $\theta_i$, $i = 1, 2$, is uniformly distribution in $[0, 2\pi)$, the joint distribution of $\theta^{*}$ and $\psi_i$, $i = 1, 2$, can be written as 
\begin{align}
	f_{\theta^*, \bm{\psi}}\left(\theta^{*}, \bm{\psi}\right) &= f_{\bm{\theta}}(\bm{\theta}) |J|\bigg|_{\theta_3=\psi_{3a}} +  f_{\bm{\theta}}(\bm{\theta}) |J|\bigg|_{\theta_3=\psi_{3b}}  \nonumber\\
	&= \frac{1}{4\pi^3}. \label{Eq-42}
\end{align}
By taking integration of \eqref{Eq-42} with respect to $\psi_i$, $i = 1, 2$, the distribution of $\theta^{*}$ can be readily derived as $f_{\Theta^{*}}(\theta^{*}) = {1}/{\pi}$, for all $\theta^{*} \in \left[\left.-\frac{\pi}{2}, \frac{\pi}{2}\right.\right)$. Accordingly, we have $f_{\Theta^*}(\theta^*) = {1}/{2\pi}$, for all $\theta^* \in [-\pi, \pi)$,  which is apparently independent of the speed distribution $v_i$. This completes the proof.

\section{Proof of Lemma~\ref{Lemma_1}}
\label{Proof_Lemma_1}
Consider the set of UAVs and the projected typical UE in Fig~\ref{Fig-7}, where the equivalent serving UAV (say, $\text{UAV}_0$) is located at $X_1$. The typical projected UE moves a distance of $v't$ in a uniformly random direction $\theta'$ from $X_1$ to $X_2$. The distance of the projected UE from the equivalent serving UAV before and after its movement is $r^*$ and $R'$, respectively. Defining $c\left(O', r^*\right)$ and $c\left(O', R' \right)$ as two circles with radius $r^*$ and $R'$, respectively, a handoff will not occur if there is no UAV in $c\left(O', R' \right)$. Since $c\left(O', r^*\right)$ is empty by definition, a handover will not occur if there is no UAV in $c\left(O', R' \right) \setminus c\left(O', r^* \right)$. Define $G$ as the event that there is no UAV in $c(O', R')$ and $\bar{H_1}(t)$ as the event that handoff has not occurred until time $t$, then $\bar{H_1}(t) \subset G$. Given $\{r^{*}, \theta', v'\}$, the conditional handoff probability can be computed as
\begin{align}
	\mathbb{P}[H_1(t)|r^*,\theta', v'] 
	&= 1- \mathbb{P}[\bar{H}_1(t)|r^*,\theta', v'] \nonumber\\
	&\geq 1-\mathbb{P}[G] \nonumber\\
	&= 1-\mathbb{P}\left[\Phi \left[ c\left(O', R'\right)\setminus c\left(O', r^* \right) \right] =0\right] \nonumber\\
	&=1-\exp\left(-2\lambda_0 |c\left(O', R\right)\setminus c\left(O', r^* \right)| \right), \label{Eq_Proof_Lemma_1}
\end{align}
where \eqref{Eq_Proof_Lemma_1} comes from the void probability of PPP. Moreover, the coverage area when the typical UE moves from $X_1$ to $X_2$ can be computed from the plane geometry by
\begin{align}
	&|c\left(O', R\right)\setminus c\left(O', r^* \right)| \nonumber\\
	&\quad = |c\left(O', R'\right)| \setminus | c\left(O', r^* \right) \cap c\left(O', R' \right)| \nonumber\\
	&\quad  = \pi R'^2 - \left({r^*}^2 \left(\phi_1 - \frac{1}{2}\sin(2\phi_1) \right) \right. \nonumber\\
	&\qquad \left. {}+R'^2 \left(\phi_2- \frac{1}{2}\sin(2\phi_2)\right) \right), \label{Eq_Proof_lemma_2}
\end{align}
where $R'$ is defined by \eqref{Eq-Distance2}, while $\phi_1$ and $\phi_2$ are defined by \eqref{Eq-21c} and \eqref{Eq-21d}, respectively. Finally, substituting \eqref{Eq_Proof_lemma_2} into \eqref{Eq_Proof_Lemma_1} yields the desired \eqref{Equation_Conditional_Handoff}.

\section{Proof of Eq.~\eqref{Equation_Handoff_1}}
\label{Proof_Eq_Handoff_1}
Starting from \eqref{Eq_Proof_Lemma_1}, the coverage area when the equivalent UAV moves from $X_1$ to $X_2$ can be computed by
\begin{equation}\label{Eq_CommomArea}
	|c(O', R')\setminus c(O',r^*)| = |c(O', R’)| -|c(O', R‘) \cap   c(O', r^*)|.
\end{equation}

\begin{figure}[t!] 
	\centering    
	\subfloat[${v’}t \leq r^*$.] 
	{
		\begin{minipage}[t]{0.25\textwidth}
			\centering         
			\includegraphics[width=0.8\textwidth]{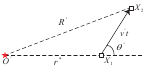}   
			\label{Fig-18a}
		\end{minipage}
	} 
	\subfloat[${v'}t > r^*$.] 
	{
		\begin{minipage}[t]{0.25\textwidth}
			\centering      
			\includegraphics[width= 0.8\textwidth]{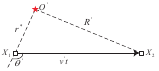}  
			\label{Fig-18b} 
		\end{minipage}
	}
	\caption{Relationships between $r^{*}$, ${v’}t$, and $\theta'$.} 
	\label{Fig-18}  
\end{figure}

Due to its symmetry, $\theta'$ is regarded as uniformly distributed within the interval $[0,\pi]$. The relationship between $r^*$ and $v't$ depicts two distinct scenarios in Fig.\ref{Fig-18}. Specifically, Fig.\ref{Fig-18a} corresponds to the scenario of $v't \leq r^*$, where $\theta' \in [0, {\pi}/{2}]$, while Fig.~\ref{Fig-18b} portrays the scenario of $v't > r^*$, with $\theta' \in ({\pi}/{2}, \pi]$.

In case $v't \leq r^*$, Fig.~\ref{Fig-18a} shows that we have $\angle X_{2} X_{1}O' = \pi - \theta’$ and $\angle O' X_2 X_{1} = \theta' - \arcsin\left({{v’}t \sin(\theta^*)}/{R'}  \right)$, then the shared area between two intersecting circles of radii $r^*$ and $R'$ can be computed as:
\begin{align}
	&{|c(O',R’) \cap   c(O',r^*)|} \nonumber\\
	& = {r^*}^2 \angle X_{2} X_{1}O' + R'^2 \angle O' X_2 X_{1} - r^*{v'}t \sin(\angle X_{2} X_{1}O'), \nonumber\\
	&= \hspace{-4pt} R’^2\left(\theta'\hspace{-4pt} - \arcsin\left({{v’}t \sin{\theta'}}/{R’}  \right)\right) + {r^*}^2(\pi - \theta') - r^* {v'}t  \sin{\theta'}.\label{Eq_CommomArea_2}
\end{align}

In case $v't > r^*$ and $\theta'$ lies within the range of $0$ to ${\pi}/{2}$ or ${v'}t \cos(\pi -\theta') \leq r^*$, Fig.~\ref{Fig-18b} shows that the angle $\angle X_{1}O'X_2= \arcsin\left({{v'}t \sin{\theta'}}/{R'}  \right)$ is constrained to the interval of $0$ to ${\pi}/{2}$. Despite this condition, \eqref{Eq_CommomArea_2} remains valid. However, in cases where $\angle X_{1}O'X_2$ falls between ${\pi}/{2}$ and $\pi$, indicating that $\theta'$ resides within the range of ${\pi}/{2}$ to $\pi$ and ${v'}t \cos(\pi -\theta') > r^*$, the term $\arcsin\left({{v'}t \sin{\theta'}}/{R'} \right)$ in \eqref{Eq_CommomArea_2} must be replaced by $\pi -\arcsin\left({{v't} \sin(\theta'}/{R'} \right)$, i.e., 
\begin{align}\label{Eq_CommomArea_3}
	|c(O', R’) \cap   c(O', r^*)| &=  R'^2\left(\pi+\theta' - \arcsin\left({{v'}t \sin{\theta'}}/{R'}  \right) \right)\nonumber\\
	&\quad {}+ r^2(\pi - \theta’)- r^* {v'}t  \sin{\theta'}.
\end{align}
Inserting \eqref{Eq_CommomArea_2}-\eqref{Eq_CommomArea_3} into \eqref{Eq_Proof_Lemma_1} and integrating with respect to $r^*$, $\theta'$, and $v’$, we attain the desired \eqref{Equation_Handoff_1}.

\section{Proof of Lemma~\ref{Lemma_2} }
\label{Proof_Lemma_2}

Before deriving the coverage probability, it is essential to characterize the probability distribution function of the channel gains. The vector $\bm{g}_i \in \mathbb{C}^{M \times 1}$ is made up of complex-valued Gaussian variables $g_k$, where $k\in \{1,2\cdots, M\}$. Each $g_k$ has the distribution $g_k \sim \mathcal{CN}(\sqrt{K}, 1)$. The squared magnitude of $g_k$, $|g_k|^2$, follows the Gamma distribution with PDF $f(x) = \left({n}/{\beta}\right)^n {x^{n-1}}\exp\left(-{n}/{\Omega}x\right)/{\Gamma(n)}$ \cite[Eq. (2.67)]{Gordon2017Principles}. Here, $n \approx (K+1)^2/(2K+1)$ and $\beta = K+1$. The channel gain $h_i$ is defined as $|\bm{g}^{H}_i\bm{\omega}_i|^2$ and approximately follows the Gamma distribution $\Gamma(n_1, \beta_1)$ with $n_1 = nM$ and $\beta_1 = \beta/n$. Based on the intrinsic properties of the Gamma distribution, it can be deduced that the variable $h_i d_i^{-\alpha}$ maintains a Gamma distribution, say, $\Gamma(n_1, \beta_1 d_i^{-\alpha})$. Regarding the interference links, the channel gain $h_j$ can be determined using Gamma distribution $\Gamma(n_2, \beta_2)$, where $n_2$ and $\beta_2$ are determined through second-order moment matching method \cite[Lemma 7]{5953530}.

 Likewise, define $Z' \triangleq \sum_{i=1}^{3}h_i d_i^{-\alpha}$, using the second-order moment matching method, we get $Z' \sim \Gamma(n', \beta')$ with 
\begin{align}
	n'=\frac{n_1 \left(\sum_{i=1}^{3}d_i^{-\alpha}\right)^2}{\sum_{i=1}^{3}\left(d_i^{-\alpha} \right)^2}, \quad \beta' =\frac{\sum_{i=1}^{3}\left(d_i^{-\alpha} \right)^2}{\beta_1 \sum_{i=1}^{3}d_i^{-\alpha}}.
\end{align}

By applying the Cauchy-Schwarz inequality, we can derive an explicit upper bound on the coverage probability given in \eqref{Eq_Theorem_2} as follows:
\begin{align}
	&\mathbb{P}_{\rm C}(t=0)= \mathbb{E}\left[ \Pr \left\{ \Upsilon > \gamma\right\} \right]  \nonumber\\
	&\quad\leq \int\limits_{0 < r_1 \leq r_2\leq r_3< \infty} \hspace{-15pt} \Pr \left\{\frac{3 \sum_{i=1}^{3}h_i d_i^{-\alpha} }{I} > \gamma |d_i\right\} \nonumber\\
	&\quad = \int\limits_{0 < r_1 \leq r_2\leq r_3< \infty} \hspace{-15pt} \Pr \left\{Z' > \frac{\gamma I}{3}   \right\} f_{r_1, r_2, r_3}({x}, y, z) \, {\rm d}x {\rm d}y {\rm d}z \nonumber\\
	&\quad = \int\limits_{0 < r_1 \leq r_2\leq r_3< \infty} \hspace{-15pt} f_{r_1, r_2, r_3}({x}, y, z) \sum_{k=0}^{\varepsilon-1}\frac{1}{k!}\left(\frac{\gamma}{3\beta'}  \right)^k \nonumber\\
	&\qquad \times \mathbb{E}_{I}\left[I^k \exp\left(-\frac{\gamma I}{3\beta'} \right)\right] \,  {\rm d}x {\rm d}y {\rm d}z 
	\label{Eq_Coverage_O_b}\\
	&\quad \leq \int\limits_{0 < r_1 \leq r_2\leq r_3< \infty} \hspace{-15pt} f_{r_{1}, r_{2}, r_{3}}(x, y, z) \sum_{k=0}^{\varepsilon-1}\frac{1}{k!}\left(- \frac{ \gamma }{3\beta_1}\right)^k \nonumber\\
	&\qquad \times  \frac{\partial^k L_{I_{1}}(s)}{\partial s^k} \bigg|_{s=\frac{ \gamma }{3 \beta'}} \,  {\rm d}x {\rm d}y {\rm d}z, 	\label{Eq_Coverage_O_c}
\end{align} 
where \eqref{Eq_Coverage_O_b} follows from the expansion of the complementary CDF of $Z$ with $\varepsilon$ defined by \eqref{Eq_varpi}, and \eqref{Eq_Coverage_O_c} comes from the relationship between the moments and Laplace transform of a random variable. Using a similar approach as in \cite[Appendix~A]{8976426}, we can derive the recurrence relation between the derivatives of $L_{I_1}(s)$, which yields \eqref{Eq_Theorem_2}.

\bibliographystyle{IEEEtran}
\bibliography{References}

\begin{IEEEbiography}
	[{\includegraphics[width=1in, height=1.25in, clip, keepaspectratio]{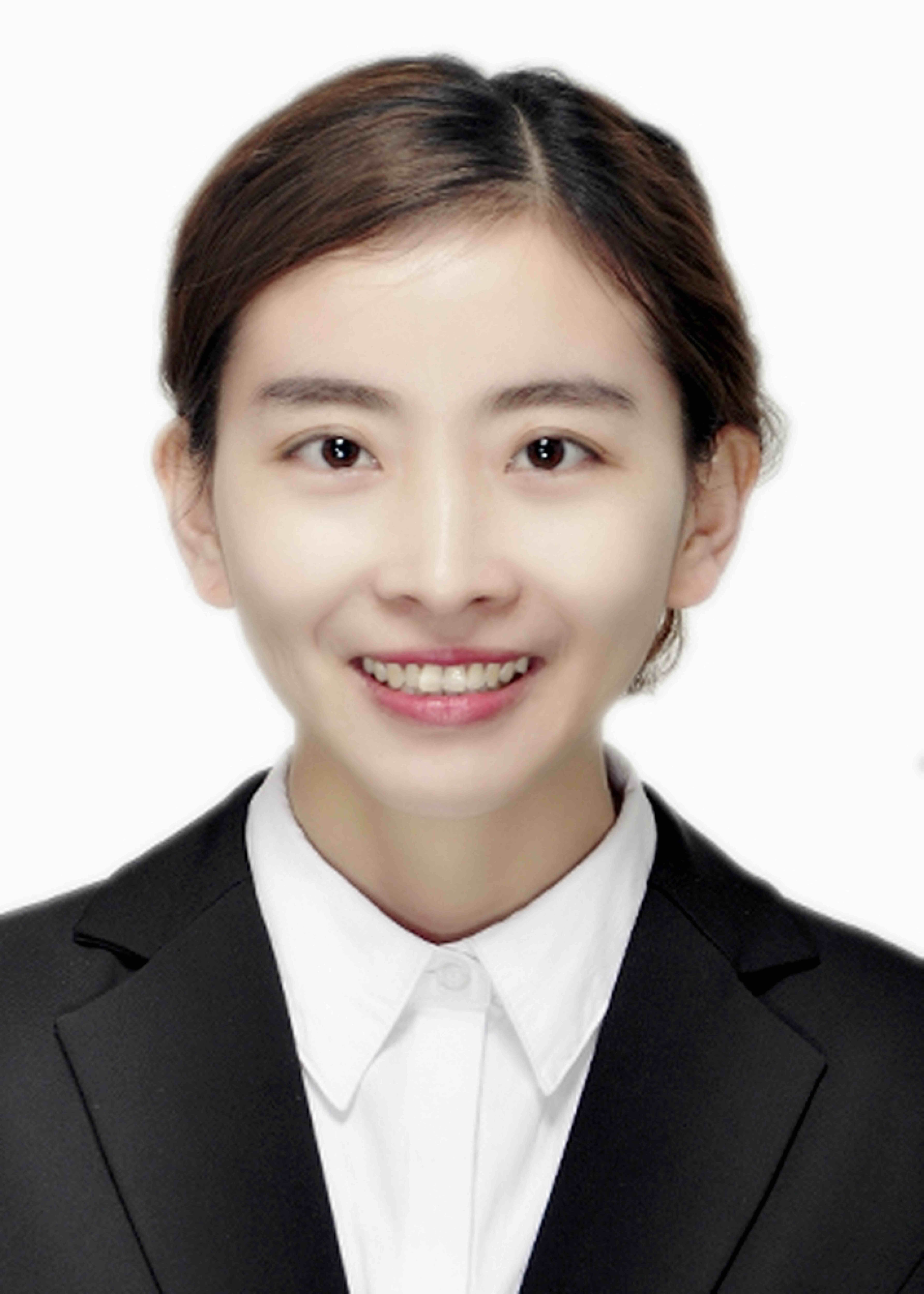}}]{Yan Li}  received a B.S. degree in Electronic Information Engineering from Hunan Normal University, Changsha, China, in 2013, and the M.S. degree in Electronics and Communication Engineering from Sun Yat-sen University, Guangzhou, China, in 2016. She received her PH.D. degree in Information and Communication Engineering at Sun Yat-sen University, Guangzhou, China, in 2021. Since 2021, she has been a Lecturer at the School of Computer and Communication Engineering, Changsha University of Science and Technology, Changsha, China. She is a Postdoctoral Fellow at the College of System Engineering, National University of Defense Technology, Changsha, China. Her research interests include modeling and analysis of UAV networks and cooperative communications based on stochastic geometry theory.
\end{IEEEbiography}

\begin{IEEEbiography}
	[{\includegraphics[width=1in, height=1.25in, clip, keepaspectratio]{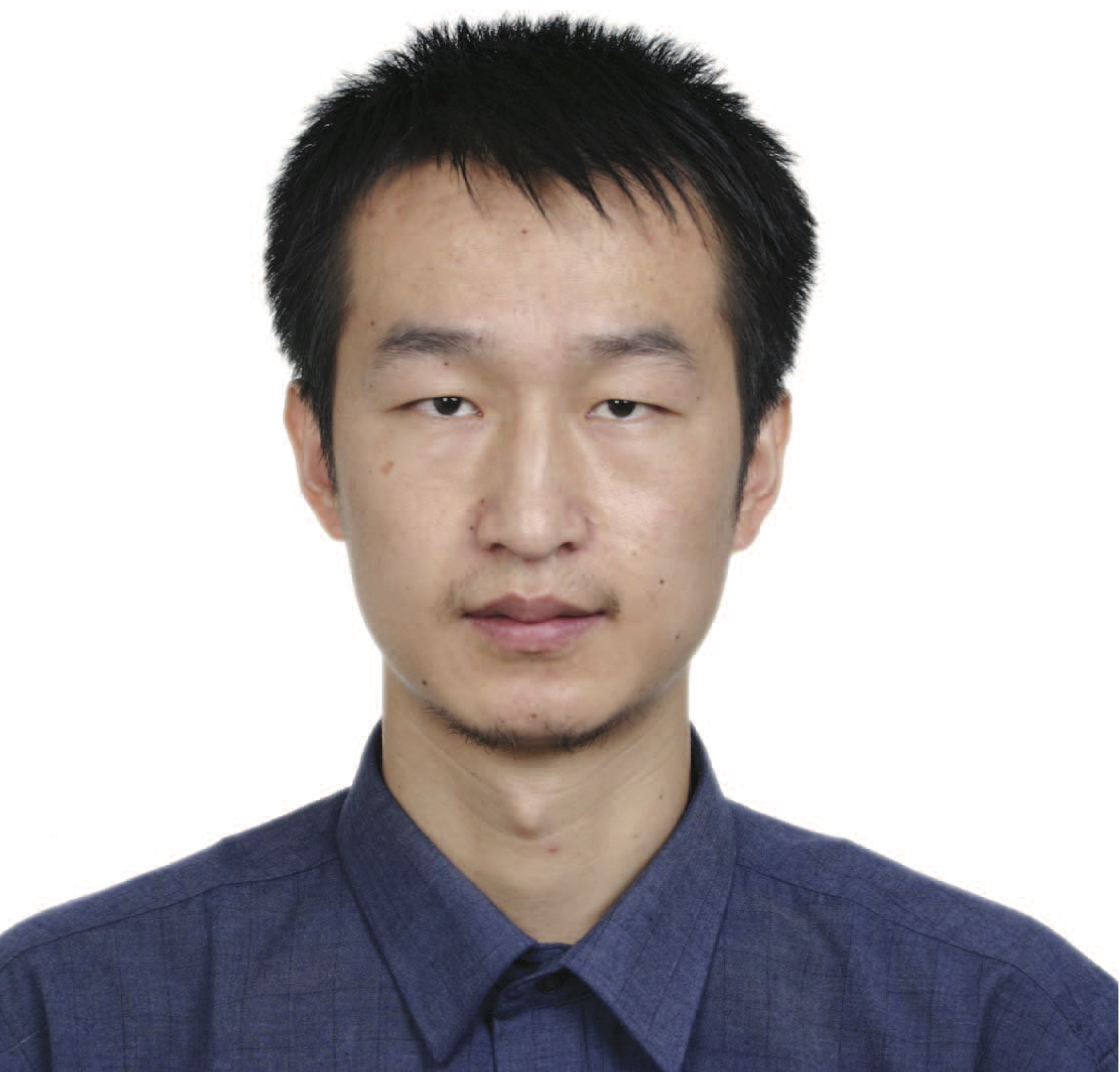}}]{Deke Guo}received the B.S. degree in industry engineering from the Beijing University of Aeronautics and Astronautics, Beijing, China, in 2001, and the Ph.D. degree in management science and engineering from the National University of Defense Technology, Changsha, China, in 2008. He is currently a Professor with the College of System Engineering, National University of Defense Technology. His research interests include distributed systems, software-defined networking, data center networking, wireless and mobile systems, and interconnection networks.
\end{IEEEbiography}

\begin{IEEEbiography}
	[{\includegraphics[width=1in, height=1.25in, clip, keepaspectratio]{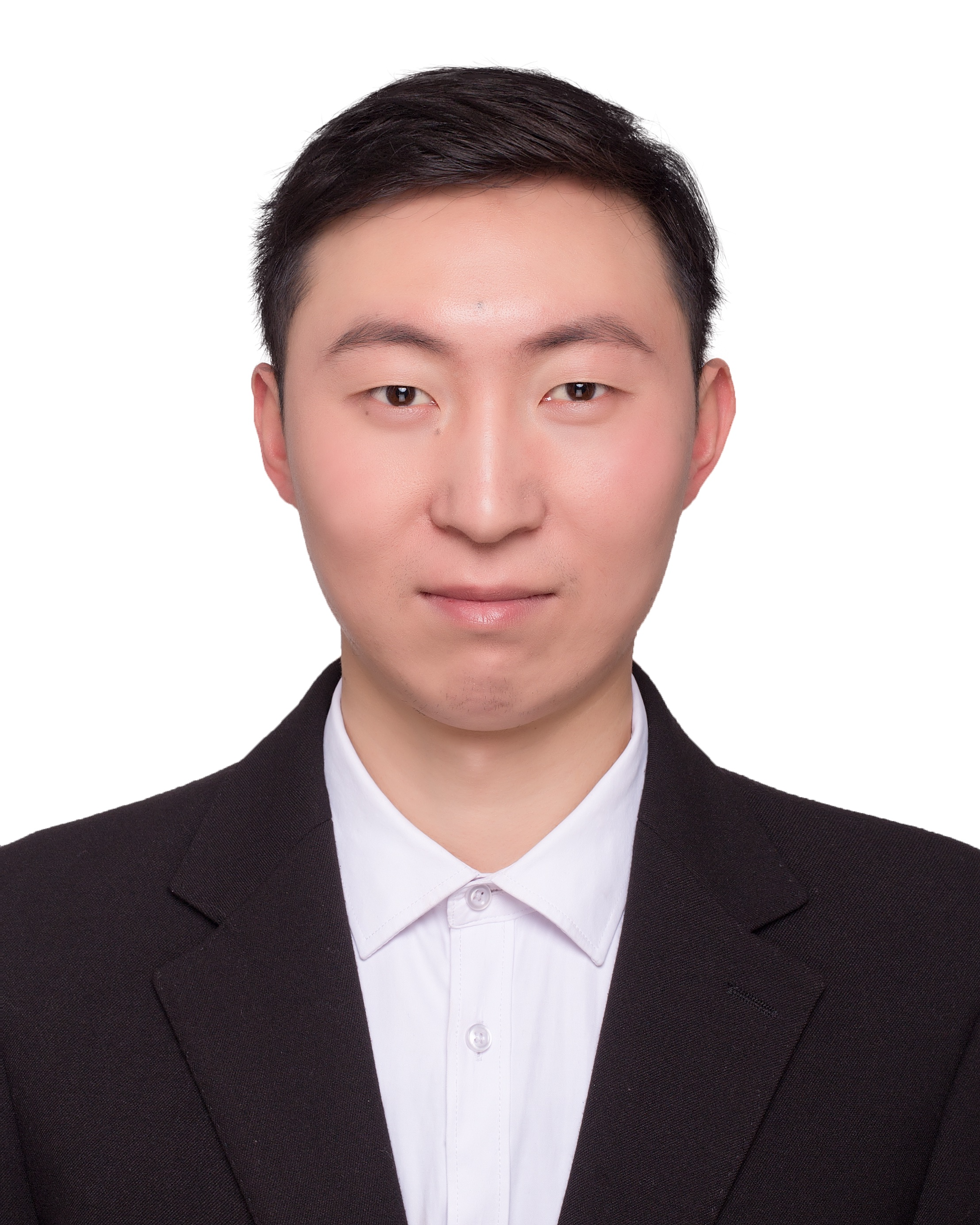}}]{Lailong Luo} received his B.S., M.S. and Ph.D degree at the College of Systems Engineering from the National University of Defence Technology, Changsha, China, in 2013, 2015, and 2019 respectively. He is currently an Associate Professor with the School of Systems, National University of Defense Technology, Changsha, China. His research interests include edge computing and distributed networking systems.
\end{IEEEbiography}

\begin{IEEEbiography}
	[{\includegraphics[width=1in, height=1.25in, clip, keepaspectratio]{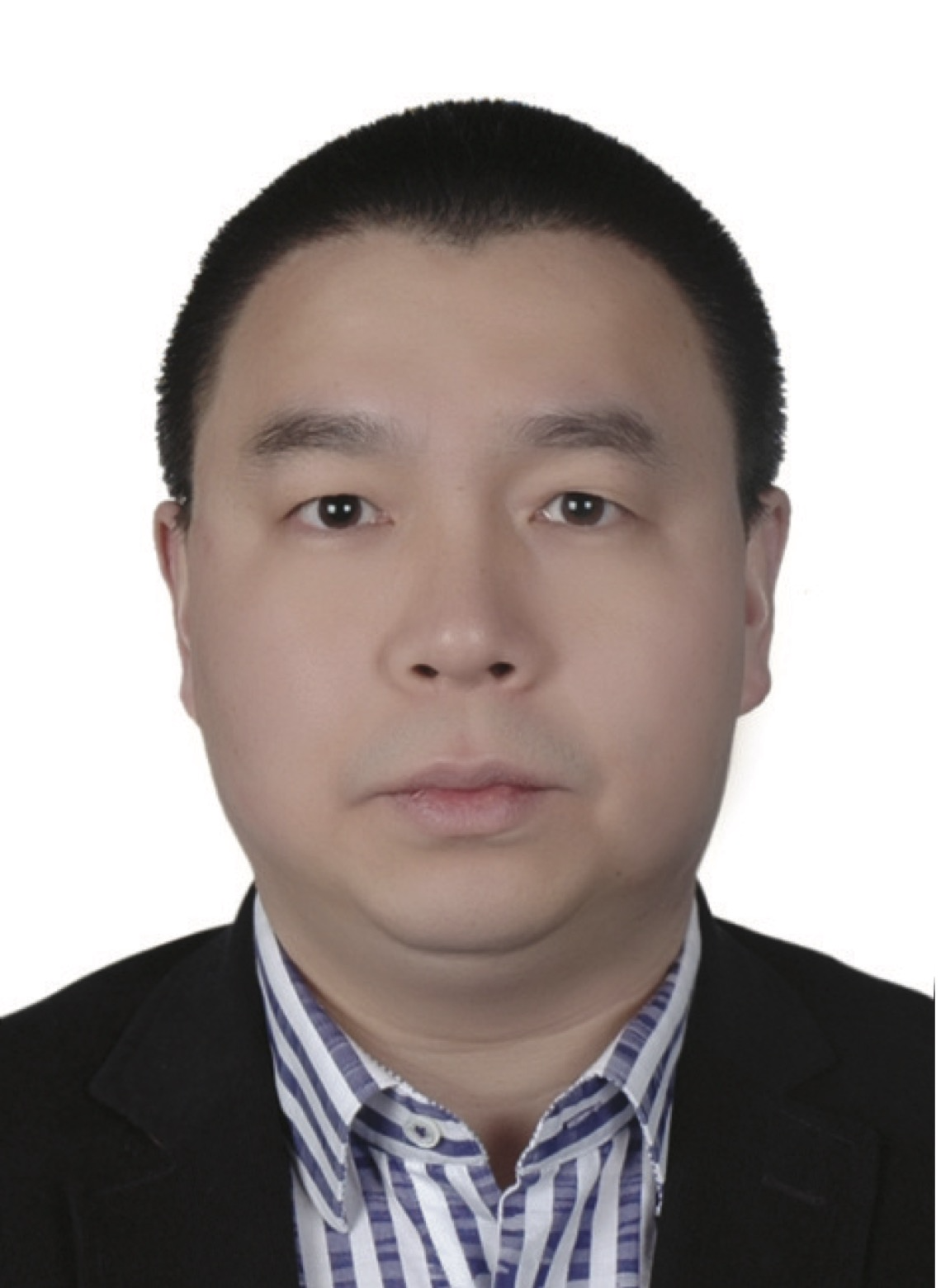}}]{Minghua Xia} (Senior Member, IEEE) received the Ph.D. degree in Telecommunications and Information Systems from Sun Yat-sen University, Guangzhou, China, in 2007. 
	
	From 2007 to 2009, he was with the Electronics and Telecommunications Research Institute (ETRI) of South Korea, Beijing R\&D Center, Beijing, China, where he worked as a member and then as a senior member of the engineering staff. From 2010 to 2014, he was in sequence with The University of Hong Kong, Hong Kong, China; King Abdullah University of Science and Technology, Jeddah, Saudi Arabia; and the Institut National de la Recherche Scientifique (INRS), University of Quebec, Montreal, Canada, as a Postdoctoral Fellow. Since 2015, he has been a Professor at Sun Yat-sen University. Since 2019, he has also been an Adjunct Professor at the Southern Marine Science and Engineering Guangdong Laboratory (Zhuhai). His research interests are in the general areas of wireless communications and signal processing. 
\end{IEEEbiography}

\end{document}